%% file: sbn.tex
\documentclass{article}

\usepackage[final]{nips_2018}

\usepackage[utf8]{inputenc} 
\usepackage[T1]{fontenc}    
\usepackage[hidelinks]{hyperref}       
\usepackage{url}            
\usepackage{booktabs}       
\usepackage{multirow}
\usepackage{amsmath}
\usepackage{amsthm}
\usepackage{amsfonts}       
\usepackage{amssymb}
\usepackage{nicefrac}       
\usepackage{microtype}      
\usepackage{graphicx}
\usepackage{breakcites}
\usepackage{tikz}
\usetikzlibrary{bayesnet}
\usepackage{bbm}
\usepackage{algorithm, algorithmic}
\usepackage{mdwlist}
\usepackage{wrapfig}
\usepackage{setspace}

\usepackage{xr}

\usepackage[skip=9pt]{caption}

\newtheorem{proposition}{Proposition}
\newtheorem{theorem}{Theorem}
\newtheorem{definition}{Definition}
\newtheorem{lemma}{Lemma}

\DeclareSymbolFont{bbold}{U}{bbold}{m}{n}
\DeclareSymbolFontAlphabet{\mathbbold}{bbold}

\title{Generalizing Tree Probability Estimation via Bayesian Networks}

\author{
  Cheng Zhang \\
  Fred Hutchinson Cancer Research Center\\
  Seattle, WA 98109 \\
  \texttt{czhang23@fredhutch.org} \\
   \And
  Frederick A. Matsen IV \\
  Fred Hutchinson Cancer Research Center\\
  Seattle WA 98109 \\
  \texttt{matsen@fredhutch.org} \\
}

\begin{document}

\maketitle

\begin{abstract}
\vspace{-0.3cm}Probability estimation is one of the fundamental tasks in statistics and machine learning.
However, standard methods for probability estimation on discrete objects do not handle object structure in a satisfactory manner.
In this paper, we derive a general Bayesian network formulation for probability estimation on leaf-labeled trees that enables flexible approximations which can generalize beyond observations.
We show that efficient algorithms for learning Bayesian networks can be easily extended to probability estimation on this challenging structured space.
Experiments on both synthetic and real data show that our methods greatly outperform the current practice of using the empirical distribution, as well as a previous effort for probability estimation on trees.
\end{abstract}

\section{Introduction}

\vspace{-0.2cm}Leaf-labeled trees, where labels are associated with the observed variables, are extensively used in probabilistic graphical models.
A typical example is the phylogenetic leaf-labeled tree, which is the fundamental structure for modeling the evolutionary history of a family of genes \citep{Felsenstein03, Friedman02}.
Inferring a phylogenetic tree based on a set of DNA sequences under a probabilistic model of nucleotide substitutions has been one of the central problems in computational biology, with a wide range of applications from genomic epidemiology \citep{Neher2015-jr} to conservation genetics \citep{DeSalle2004-mm}.
To account for the phylogenetic uncertainty, Bayesian approaches are adopted \citep{Huelsenbeck01b} and Markov chain Monte Carlo (MCMC) \citep{Yang97, Mau99, Huelsenbeck01} is commonly used to sample from the posterior of phylogenetic trees.
Posterior probabilities of phylogenetic trees are then typically estimated with simple sample relative frequencies (SRF), based on those MCMC samples.

While classical, this empirical approach is unsatisfactory for tree posterior estimation due to the combinatorially exploding size of tree space.
Specifically, SRF does not support trees beyond observed samples (i.e., simply sets the probabilities of unsampled trees to zero), and is prone to unstable estimates for low-probability trees.
As a result, reliable estimations using SRF usually require impractically large sample sizes.
Previous work \citep{Hhna2012-pm, Larget2013-et} attempted to remedy these problems by harnessing the similarity among trees and proposed several probability estimators using MCMC samples based on conditional independence of separated subtrees.
Although these estimators do extend to unsampled trees, the conditional independence assumption therein is often too strong to provide accurate approximations for posteriors inferred from real data \citep{Whidden2015-eq}.

In this paper, we present a general framework for tree probability estimation given a collection of trees (e.g., MCMC samples) by introducing a novel structure called \emph{subsplit Bayesian networks} (SBNs).
This structure provides rich distributions over the entire tree space and hence differs from existing applications of Bayesian networks in phylogenetic inference \citep[e.g.][]{Strimmer2000,Hohna14} to compute tree likelihood.
Moreover, SBNs relax the conditional clade independence assumption and allow easy adjustment for a variety of flexible dependence structures between subtrees.
They also allow many efficient learning algorithms for Bayesian networks to be easily extended to tree probability estimation.
Inspired by weight sharing used in convolutional neural networks \citep{LeCun98}, we propose \emph{conditional probability sharing} for learning SBNs, which greatly reduces the number of free parameters by exploiting the similarity of local structures among trees.
Although initially proposed for rooted trees, we show that SBNs can be naturally generalized to unrooted trees, which leads to a missing data problem that can be efficiently solved through expectation maximization.
Finally, we demonstrate that SBN estimators greatly outperform other tree probability estimators on both synthetic data and a benchmark of challenging phylogenetic posterior estimation problems.
The SBN framework also works for general leaf-labeled trees, however for ease of presentation, we restrict to leaf-labeled bifurcating trees in this paper.

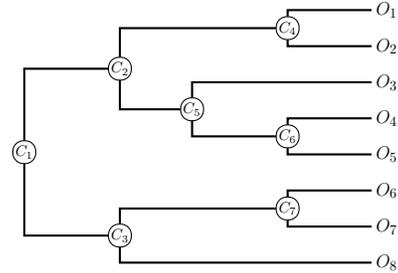
\begin{wrapfigure}[14]{r}{.4\textwidth}
\vspace{-1.4cm}
\begin{center}
\input{figs/clade-decomp-square-edge}
\caption{A rooted tree and its clade decomposition. Each clade corresponds to the set of offspring leaves (e.g., $C_2=\{O_1, O_2, O_3, O_4, O_5\} \text{ and } C_3=\{O_6, O_7, O_8\}$). }\label{fig:tree}
\end{center}
\end{wrapfigure}

\section{Background}
A leaf-labeled bifurcating tree is a binary tree (rooted or unrooted) with labeled leaves (e.g., leaf nodes associated with observed variables or a set of labels); we refer to it as a tree for short.
Recently, several probability estimators on tree spaces have been proposed that exploit the similarity of \emph{clades}, a local structure of trees, to generalize beyond observed trees.
Let $\mathcal{X}=\{O_1, \ldots, O_N\}$ be a set of $N$ labeled leaves.
A \emph{clade} $X$ of $\mathcal{X}$ is a nonempty subset of $\mathcal{X}$.
Given a rooted tree $T$ on $\mathcal{X}$, one can find its unique clade decomposition as follows.
Start from the root, which has a trivial clade that contains all the leaves $C_1$.
This clade first splits into two subclades $C_2,\;C_3$.
The splitting process continues recursively onto each successive subclade until there are no subclades to split.
Finally, we obtain a collection of nontrivial clades $T_{\mathcal{C}}$. As for the tree in Figure~\ref{fig:tree}, $T_{\mathcal{C}} = \{C_2, C_3, C_4, C_5, C_6, C_7\}$.
This way, $T$ is represented uniquely as a set of clades $T_{\mathcal{C}}$.
Therefore, distributions over the tree space can be specified as distributions over the space of sets of clades. Again, for Figure~\ref{fig:tree}:
\begin{equation}\label{eq:splits}
p(T) = p(T_{\mathcal{C}})=p(C_2, C_3, C_4, C_5, C_6, C_7)
\end{equation}
The clade decomposition representation enables distributions that reflect the similarity of trees through the local clade structure.
However, a full parameterization of this approach over all rooted trees on $\mathcal{X}$ using rules of conditional probability is intractable even for a moderate $N$.
\citet{Larget2013-et}, building on work of \citet{Hhna2012-pm}, introduced the \emph{Conditional Clade Distribution} (CCD) which assumes that given the existence of an edge in a tree, clades that further refine opposite sides of the edge are independent. CCD greatly reduces the number of parameters. For example, \eqref{eq:splits} has the following CCD approximation
\[
p_{\mathrm{ccd}}(T) = p(C_2, C_3) p(C_4, C_5|C_2) p(C_6|C_5) p(C_7|C_3)
\]
However, CCD also introduces strong bias which makes it insufficient to capture the complexity of inferred posterior distributions on real data (see Figure \ref{fig:ds1}).
In particular, certain clades may depend on their sisters.
This motivates a more flexible set of approximate distributions.

\section{A Subsplit Bayesian Network Formulation}
\vspace{-0.2cm}In addition to the clade decomposition representation, a rooted tree $T$ can also be uniquely represented as a set of subsplits.
Let $\succ$ be a total order on clades (e.g., lexicographical order).
A \emph{subsplit} $(Y, Z)$ of a clade $X$ is an ordered pair of disjoint subclades of $X$ such that $Y\cup Z=X$ and $Y\succ Z$.
For example, the tree in Figure \ref{fig:tree} corresponds to the following set of nontrivial subsplits
\[
T_{\mathcal{S}} = \{(C_2,C_3), (C_4,C_5), (\{O_3\},C_6), (C_7,\{O_8\})\}
\]
with lexicographical order on clades. Moreover, this set-of-subsplits representation of trees inspires a natural probabilistic Bayesian network formulation as follows (Figure~\ref{fig:bn}):

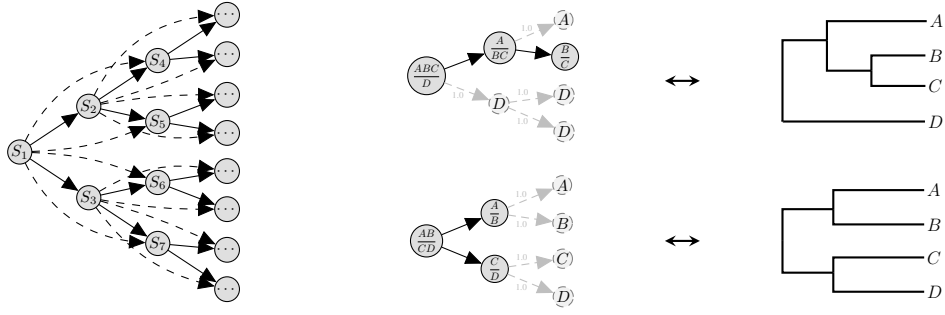
\begin{figure*}[t!]
\begin{center}
\input{figs/bn-reversed.tex}
\caption{The subsplit Bayesian network formulation.
{\bf Left}: A general Bayes net for rooted trees.
Each node represents a subsplit-valued or singleton-clade-valued random variable.
The solid full and complete binary tree network is $\mathcal{B}^\ast_\mathcal{X}$.
{\bf Middle/Right}: Examples of rooted trees with 4 leaves.
Note that the \emph{solid dark nets} that represent the true splitting processes of the trees may have dynamical structures.
By allowing singleton clades continue to split (the \emph{dashed gray nets}) until depth of 3, both nets grow into the full and complete binary tree of depth 3.}\label{fig:bn}
\end{center}
\vspace{-1.5em}
\end{figure*}

\begin{definition}
A \emph{subsplit Bayesian network (SBN)} $\mathcal{B}_\mathcal{X}$ on a leaf set $\mathcal{X}$ of size $N$ is a Bayesian network whose nodes take on subsplit values or singleton clade values of $\mathcal{X}$ and: i) has depth $N-1$ (the root counts as depth 1); ii) The root node takes on subsplits of the entire leaf set $\mathcal{X}$; iii) contains a full and complete binary tree network $\mathcal{B}^\ast_\mathcal{X}$ as a subnetwork.
\end{definition}

Note that $\mathcal{B}^\ast_\mathcal{X}$ itself is an SBN and is contained in all SBNs; therefore, $\mathcal{B}^\ast_\mathcal{X}$ is the minimum SBN on $\mathcal{X}$.
Moreover, $\mathcal{B}^\ast_\mathcal{X}$ induces a natural indexing procedure for the nodes of all SBNs on $\mathcal{X}$: starting from the root node, which is denoted as $S_1$, for any $i$, we denote the two children of $S_i$ as $S_{2i}$ and $S_{2i+1}$, respectively, until a leaf node is reached.
We call the parent nodes in $\mathcal{B}^\ast_\mathcal{X}$ the \emph{natural parents}.

\begin{definition}\label{def:compatible}
We say a subsplit $(Y,Z)$ is \emph{compatible} with a clade $X$ if $Y\cup Z = X$.
Moreover, a singleton clade $\{W\}$ is said to be compatible with itself.
With natural indexing, we say a full SBN assignment $\{S_i=s_i\}_{i\ge1}$ is \emph{compatible} if for any interior node assignment $s_i=(Y_i, Z_i)$ (or $\{W_i\}$), $s_{2i}, s_{2i+1}$ are compatible with $Y_i, Z_i$ (or $\{W_i\}$), respectively.
Consider a parent-child pair in an SBN, $S_i$ and $S_{\pi_i}$, where $\pi_i$ denotes the index set of the parent nodes of $S_i$.
We say an assignment $S_i=s_i, S_{\pi_i}=s_{\pi_i}$ is \emph{compatible} if it can be extended to a compatible assignment of the SBN.
\end{definition}

\begin{lemma}\label{lemma:sbn}
Given an SBN $\mathcal{B}_\mathcal{X}$, each rooted tree $T$ on $\mathcal{X}$ can be uniquely represented as a compatible assignment of $\mathcal{B}_\mathcal{X}$.
\end{lemma}

A proof of Lemma \ref{lemma:sbn} is provided in the Appendix.
Unlike in phylogenies, nodes in SBNs take on subsplit (or singleton clade) values that represent the local topological structure of trees.
By including the true splitting processes (e.g., $T_\mathcal{S}$) of the trees while allowing singleton clades continue to split (``fake'' split) until the whole network reaches depth $N-1$ (see Figure~\ref{fig:bn}), each SBN on $\mathcal{X}$ has a fixed structure which contains the full and complete binary tree as a subnetwork.
Note that those fake splits are all deterministically assigned, which means the corresponding conditional probabilities are all one.
Therefore, the estimated probabilities of rooted trees only depend on their true splitting processes.
With SBNs, we can easily construct a family of flexible approximate distributions on the tree space.
For example, using the minimum SBN, \eqref{eq:splits} can be estimated as
\[
p(C_2, C_3) p(C_4, C_5|C_2, C_3) p(C_6|C_4, C_5) p(C_7|C_2,C_3)
\]
This approximation implicitly assumes that given the existence of a subsplit, subsplits that further refine opposite sides of this subsplit are independent.
Note that CCD can be viewed as a simplification where the conditional probabilities are further approximated as follows
\[
p(C_4, C_5 |C_2, C_3) \approx p(C_4, C_5|C_2),\quad p(C_6|C_4, C_5) \approx p(C_6|C_5), \quad p(C_7|C_2, C_3) \approx p(C_7|C_3)
\]
By including the sister clades in the conditional subsplit probabilities, SBNs relax the conditional clade independence assumption made in CCD and allows for more flexible dependent structures between local components (e.g., subsplits in sister-clades).
Moreover, one can add more complicated dependencies between nodes (e.g., dashed arrows in Figure \ref{fig:bn}(a)) and hence easily adjust SBN formulation to provide a wide range of flexible approximate distributions. For general SBNs, the estimated tree probabilities take the following form:
\begin{equation}\label{eq:sbn}
p_{\mathrm{sbn}}(T) = p(S_1)\prod_{i>1}p(S_i|S_{\pi_{i}}).
\end{equation}

In addition to the superior flexibility, another benefit of the SBN formulation is that these approximate distributions are all naturally normalized if the conditional probability distributions (CPDs) are \emph{consistent}, as defined next:
\begin{definition}
We say the conditional probability $p(S_i|S_{\pi_i})$ is \emph{consistent} if $p(S_i=s_i|S_{\pi_i}=s_{\pi_i}) = 0$ for any incompatible assignment $S_i=s_i, S_{\pi_i}=s_{\pi_i}$.
\end{definition}
\begin{proposition}\label{prop:rooted}
If $\forall i>1$, $p(S_i|S_{\pi_i})$ is \emph{consistent}, then $\sum_{T} p_{\mathrm{sbn}}(T) = 1$.
\end{proposition}
With Lemma~\ref{lemma:sbn}, the proof is standard and is given in the Appendix.
Furthermore, the SBN formulation also allows us to easily extend many efficient algorithms for learning Bayesian networks to SBNs for tree probability estimation, as we see next.

\section{Learning Subsplit Bayesian Networks}
\vspace{-0.2cm}
\subsection{Rooted Trees}
\vspace{-0.2cm}Suppose we have a sample of rooted trees $\mathcal{D}=\{T_k\}_{k=1}^K$ (e.g., from a phylogenetic MCMC run given DNA sequences).
As before, each sampled tree can be represented as a collection of subsplits $T_k=\{S_{i}=s_{i,k},\;i\geq 1\},\; k=1,\ldots,K$
and therefore has the following SBN likelihood
\[
L(T_k) = p(S_1=s_{1,k})\prod_{i>1}p(S_i=s_{i,k}|S_{\pi_i}=s_{\pi_i,k}).
\]

\paragraph{Maximum Likelihood}
\vspace{-.3cm}In this complete data scenario, we can simply use maximum likelihood to learn the parameters of SBNs. Denote the set of all observed subsplits of node $S_i$ as $\mathbb{C}_i, \; i\geq1$ and the set of all observed subsplits of the parent nodes of $S_i$ as $\mathbb{C}_{\pi_i},\; i>1$.
Assuming that trees are independently sampled, the complete data log-likelihood is
\begin{align}
\log L(\mathcal{D}) &=\sum_{k=1}^K \bigg(\log p(S_1=s_{1,k}) + \sum_{i>1} \log p(S_{i}=s_{i,k}|S_{\pi_i}=s_{\pi_i,k})\bigg)\nonumber\\
   &=\sum_{s_1\in\mathbb{C}_1}m_{s_1}\log p(S_1=s_1) + \sum_{i>1}\sum_{\substack{s_{i}\in\mathbb{C}_i \\ t_i\in\mathbb{C}_{\pi_i}}}m_{s_i,t_i}\log p(S_{i}=s_i|S_{\pi_i}=t_i)
\label{eq:logll}
\end{align}
where $m_{s_1}=\sum_{k=1}^K\mathbbold{I}({s_{1,k}=s_1}),\; m_{s_i,t_i}=\sum_{k=1}^K\mathbbold{I}({s_{i,k}=s_i, s_{\pi_i,k}=t_i}), \; i>1$ are the frequency counts of the root splits and parent-child subsplit pairs for each interior node respectively, and $\mathbbold{I}(\cdot)$ is the indicator function.
The maximum likelihood estimates of CPDs have the following simple closed form expressions in terms of relative frequencies:
\[
\hat{p}^{\mathrm{ML}}(S_1=s_1) = \frac{m_{s_1}}{\sum_{s\in\mathbb{C}_1}m_{s}} = \frac{m_{s_1}}{K},\quad
\hat{p}^{\mathrm{ML}}(S_i=s_i|S_{\pi_i}=t_i) = \frac{m_{s_i,t_i}}{\sum_{s\in\mathbb{C}_i}m_{s,t_i}}.
\]

\paragraph{Conditional Probability Sharing}
\vspace{-.2cm}We can use the similarity of local structures to further reduce the number of SBN parameters and achieve better generalization, similar to weight sharing for convolutional nets.
Indeed, different trees do share lots of local structures, such as subsplits and clades.
 As a result, the representations of the trees in an SBN could have the same parent-child subsplit pairs, taken by different nodes (see Figure \ref{fig:CPS} in the Appendix).
Instead of assigning independent sets of parameters for those pairs at different locations, we can use one set of parameters for each of those shared pairs, regardless of their locations in SBNs.
We call this specific setting of parameters in SBNs \emph{conditional probability sharing}. Compared to standard Bayes nets, this index-free parameterization only needs CPDs for each observed parent-child subsplit pair, dramatically reducing the number of parameters in the model.


Now denote the set of all observed splits of $S_1$ as $\mathbb{C}_r$, and the set of all observed parent-child subsplit pairs as $\mathbb{C}_{\mathrm{ch}|\mathrm{pa}}$.
The log-likelihood $\log L(\mathcal{D})$ in \eqref{eq:logll} can be rewritten into
\[
\log L(\mathcal{D})=\sum_{s_1\in\mathbb{C}_r}\!\!m_{s_1}\log p(S_1=s_1) + \!\!\!\!\sum_{s|t \in \mathbb{C}_{\mathrm{ch}|\mathrm{pa}}}\!\!\!\!\!\!m_{s,t}\log p(s|t)
\]
where $m_{s,t} = \sum_{k=1}^K\sum_{i>1}\mathbbold{I}({s_{i,k}=s, s_{\pi_i,k}=t})$ is the frequency count of the corresponding subsplit pair $s|t\in \mathbb{C}_{\mathrm{ch}|\mathrm{pa}}$.
Similarly, we have the maximum likelihood estimates of the CPDs for those parent-child subsplit pairs:
\begin{equation}\label{eq:dl}
\hat{p}^{\mathrm{ML}}(s|t) = \dfrac{m_{s,t}}{\sum_s m_{s,t}} ,\quad s|t \in \mathbb{C}_{\mathrm{ch}|\mathrm{pa}}.
\end{equation}
The main computation is devoted to the collection of the frequency counts $m_{s_1}$ and $m_{s,t}$ which requires iterating over all sampled trees and for each tree, looping over all the edges.
Thus the overall computational complexity is $\mathcal{O}(KN)$.

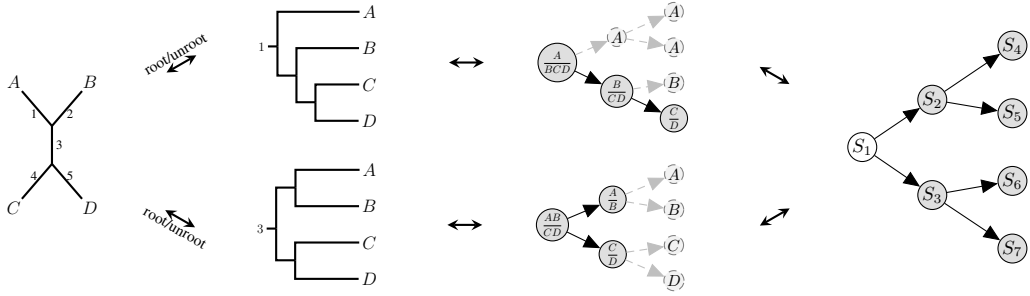
\begin{figure}[t!]
\begin{center}
\input{figs/bn-unrooted}
\caption{SBNs for unrooted trees.
{\bf Left}: A simple four taxon unrooted tree example.
It has five edges, $1,2,3,4,5$, that can be rooted on to make (compatible) rooted trees.
{\bf Middle(left)}: Two rooted trees when rooting on edges $1$ and $3$.
{\bf Middle(right)}: The corresponding SBN representations for the two rooted trees.
{\bf Right}: An integrated SBN for the unrooted tree with unobserved root node $S_1$.}
\label{fig:sbnunrooted}
\end{center}
\vskip -1.5em
\end{figure}

\subsection{Unrooted Trees}
\vspace{-0.2cm}Unrooted trees are commonly used to express undirected relationships between observed variables, and are the most common tree type in phylogenetics.
The SBN framework can be easily generalized to unrooted trees because each unrooted tree can be transformed into a rooted tree by placing the root on one of its edges.
Since there are multiple possible root placements, each unrooted tree has multiple representations in terms of SBN assignments for the corresponding rooted trees.
Unrooted trees, therefore, can be represented using the same SBNs for rooted trees, with root node $S_1$ being unobserved\footnote{The subsplits $S_2,S_3,\ldots$ are well defined once the split in $S_1$ (or equivalently the root) is given.} (Figure \ref{fig:sbnunrooted}).
Marginalizing out the unobserved node $S_1$, we obtain SBN probability estimates for unrooted trees (denoted by $T^\mathrm{u}$ in the sequel):
\begin{equation}\label{eq:unrootedprob}
p_{\mathrm{sbn}}(T^\mathrm{u}) = \sum_{S_1\sim T^{\mathrm{u}}}p(S_1)\prod_{i>1}p(S_i|S_{\pi_i})
\end{equation}
where $\sim$ means all root splits that are compatible with $T^{\mathrm{u}}$.
Similar to the SBN approximations for the rooted trees, \eqref{eq:unrootedprob} provides a natural probability distribution over unrooted trees (see a proof in the Appendix).

\begin{proposition}\label{prop:unrooted}
Suppose that the conditional probability distributions $p(S_i|S_{\pi_i}),\; i>1$ are consistent, then \eqref{eq:unrootedprob} is a probability distribution over unrooted trees with leaf set $\mathcal{X}$, that is, $\sum_{T^\mathrm{u}} p_{\mathrm{sbn}}(T^\mathrm{u}) = 1$.
\end{proposition}

\vspace{-0.1cm}As before, assume that we have a sample of unrooted trees $\mathcal{D}^\mathrm{u}=\{T_k^\mathrm{u}\}_{k=1}^K$.
Each pair of the unrooted tree and rooting edge corresponds to a rooted tree that can be represented as: $
(T_k^\mathrm{u}, e) = \{S_{i} = s_{i,k}^e, i\geq 1\}, \; e \in E(T_k^\mathrm{u}), \; 1\leq k\leq K
$
where $E(T_k^\mathrm{u})$ denotes the edges of $T_k^\mathrm{u}$ and $s_{i,k}^e$ denotes all the resulting subsplits when $T_k^\mathrm{u}$ is rooted on edge $e$.
The SBN likelihood for the unrooted tree $T_k^\mathrm{u}$ is
\[
L(T_k^\mathrm{u}) = \!\!\sum_{e\in E(T_k^\mathrm{u})} \!\!p(S_1=s_{1,k}^e)\prod_{i>1}p(S_i=s_{i,k}^e|S_{\pi_i}=s_{\pi_i,k}^e).
\]
The lost information on the root node $S_1$ means the SBN likelihood for unrooted trees can no longer be factorized.
We, therefore, propose the following two algorithms to handle this challenge.

\paragraph{Maximum Lower Bound Estimates}
\vspace{-0.2cm}A simple strategy is to construct tractable lower bounds via variational approximations \citep{Wainwright08}
\begin{equation}\label{eq:lb}
LB_q(T^\mathrm{u}) := \sum_{S_1\sim T^{\mathrm{u}}}q(S_1) \left(\log \frac{p(S_1)\prod_{i>1} p(S_i|S_{\pi_i})}{q(S_1)}\right) \le \log L(T^\mathrm{u})
\end{equation}
where $q$ is a probability distribution on $S_1\sim T^\mathrm{u}$.
In particular, taking $q$ to be uniform on the $2N-3$ tree edges together with conditional probability sharing gives the \emph{simple average} (SA) lower bound of the data log-likelihood
\[
LB^{\mathrm{SA}}(\mathcal{D}^\mathrm{u}) := \bigg(\sum_{s_1\in\mathbb{C}_r} m^\mathrm{u}_{s_1}  \log p(S_1=s_1)
+  \sum_{s|t\in\mathbb{C}_{\mathrm{ch}|\mathrm{pa}}} m_{s,t}^\mathrm{u}\log p(s|t)\bigg) + K\log(2N-3)
\]
where
\[
m^\mathrm{u}_{s_1} = \sum_{k=1}^K\sum_{e\in E(T_k^{\mathrm{u}})} \frac{1}{2N-3}\mathbbold{I}({s_{1,k}^e=s_1}),\quad
m^\mathrm{u}_{s,t} = \sum_{k=1}^K\sum_{e\in E(T_k^{\mathrm{u}})}\frac{1}{2N-3}\sum_{i>1} \mathbbold{I}({s_{i,k}^e=s, s_{\pi_i,k}^e = t}).
\]
The maximum SA lower bound estimates are then
\[
\hat{p}^{\mathrm{SA}}(S_1=s_1) = \frac{m_{s_1}^\mathrm{u}}{\sum_{s\in\mathbb{C}_r} m_s^\mathrm{u}} = \frac{m_{s_1}^\mathrm{u}}{K},\quad s_1\in \mathbb{C}_r,\quad
\hat{p}^\mathrm{SA}(s|t) = \frac{m_{s,t}^\mathrm{u}}{\sum_s m_{s,t}^\mathrm{u}},\quad s|t \in \mathbb{C}_{\mathrm{ch}|\mathrm{pa}}.
\]

\paragraph{Expectation Maximization}
\vspace{-.1cm}The maximum lower bound approximations can be improved upon by adapting the variational distribution $q$, instead of using a fixed one. This, together with conditional probability sharing, leads to an extension of the expectation maximization (EM) algorithm for learning SBNs, which also allows us use of the Bayesian formulation.
Specifically, at the E-step of the $n$-th iteration, an adaptive lower bound is constructed through \eqref{eq:lb} using the conditional probabilities of the missing root node
\[
q^{(n)}_k(S_1) = p(S_1|T^\mathrm{u}_k, \hat{p}^{\mathrm{EM},(n)}),\quad k=1,\ldots,K
\]
given $\hat{p}^{\mathrm{EM},(n)}$, the current estimate of the CPDs.
The lower bound contains a constant term that only depends on the current estimates, and a score function for the CPDs $p$
\[
Q^{(n)}(\mathcal{D}^\mathrm{u}; p) = \sum_{k=1}^K Q^{(n)}(T^\mathrm{u}_k;p) = \sum_{k=1}^K\sum_{S_1\sim T^\mathrm{u}_k} q^{(n)}_k(S_1) \big(\log p(S_1) + \sum_{i>1} \log p(S_i|S_{\pi_i})\big)
\]
which is then optimized at the M-step.
This variational perspective of the EM algorithm was found and discussed by \citet{Neal98}.
The following theorem guarantees that maximizing (or improving) the $Q$ score is sufficient to improve the objective likelihood.

\begin{theorem}\label{thm:em}
Let $T^\mathrm{u}$ be an unrooted tree. $\forall p$,
\[
Q^{(n)}(T^\mathrm{u};p) - Q^{(n)}(T^\mathrm{u}; \hat{p}^{\mathrm{EM},(n)}) \leq \log L(T^\mathrm{u};p) - \log L(T^\mathrm{u}; \hat{p}^{\mathrm{EM},(n)}).
\]
\end{theorem}

\begin{algorithm}[tb]
\setstretch{0.9}
   \caption{Expectation Maximization for SBN}
   \label{alg:SBNEM}
\begin{algorithmic}
   \STATE {\bfseries Input:} Data $\mathcal{D}^\mathrm{u}=\{T^\mathrm{u}_k\}_{k=1}^K$, regularization coeff $\alpha$.
   \STATE Initialize $\hat{p}^{\mathrm{EM},(0)}$ (e.g., via $\hat{p}^\mathrm{SA}$) and $n=0$. Set equivalent counts $\tilde{m}_{s_1}^\mathrm{u}, \tilde{m}_{s,t}^\mathrm{u}$ for regularization.
   \REPEAT
   \STATE {\bfseries E-step.} $\forall \; 1\le k \le K$, compute $
q^{(n)}_k(S_1)  = \frac{p(T_k^\mathrm{u}, S_1|\hat{p}^{\mathrm{EM},(n)})}{\sum_{S_1\sim T_k^\mathrm{u}}p(T_k^\mathrm{u}, S_1|\hat{p}^{\mathrm{EM},(n)})}
$
   \STATE {\bfseries M-step.} Compute the expected frequency counts with conditional probability sharing  and update the CPDs by maximizing the regularized $Q$ score
      {\scriptsize \begin{align*}
      \hat{p}^{\mathrm{EM},(n+1)}(S_1=s_1)  = \frac{\overline{m}_{s_1}^{\mathrm{u},(n)}+\alpha \tilde{m}_{s_1}^\mathrm{u}}{K+\alpha\sum_{s_1\in \mathbb{C}_r}\tilde{m}_{s_1}^\mathrm{u}},\;s_1\in \mathbb{C}_r,\quad \overline{m}_{s_1}^\mathrm{u, (n)}&=\sum_{k=1}^K\sum_{e\in E(T_k^\mathrm{u})} q^{(n)}_k(S_1=s_1)  \mathbbold{I}({s_{i,k}^e=s_1})\\[-5pt]
      \hat{p}^{\mathrm{EM},(n+1)}(s|t) = \frac{\overline{m}_{s,t}^{\mathrm{u},(n)}+\alpha \tilde{m}_{s,t}^\mathrm{u}}{\sum_s (\overline{m}_{s,t}^{\mathrm{u},(n)}+\alpha \tilde{m}_{s,t}^\mathrm{u})},\; s|t \in \mathbb{C}_{\mathrm{ch}|\mathrm{pa}},\quad \overline{m}^{\mathrm{u},(n)}_{s,t} \!=\! \sum_{k=1}^K&\sum_{e\in E(T_k^{\mathrm{u}})}\!\!\!\!q^{(n)}_k(S_1=s^e_{1,k})\!\sum_{i>1} \mathbbold{I}({s_{i,k}^e=s, s_{\pi_i,k}^e = t})
   \end{align*}
   \vskip-2em}
   \STATE $n \gets n+1$
   \UNTIL{convergence.}
\end{algorithmic}
\end{algorithm}

When data is insufficient or the number of parameters is large, the EM approach also easily incorporates regularization \citep{dempster77}.
Taking conjugate Dirichlet priors \citep{Buntine91}, the regularized score function is
\[
Q^{(n)}(\mathcal{D}^\mathrm{u}; p) + \sum_{s_1\in\mathbb{C}_r}\alpha\tilde{m}_{s_1}^\mathrm{u} \log p(S_1=s_1) + \sum_{s|t\in \mathbb{C}_{\mathrm{ch}|\mathrm{pa}}}\alpha\tilde{m}_{s,t}^\mathrm{u}\log p(s|t)
\]
where $\tilde{m}_{s_1}^\mathrm{u}, \tilde{m}_{s,t}^\mathrm{u}$ are the equivalent sample counts and $\alpha$ is the global regularization coefficient.
We then simply maximize the regularized score in the same manner at the M-step.
Similarly, this guarantees that the regularized log-likelihood is increasing at each iteration.
We summarize the EM approach in Algorithm \ref{alg:SBNEM}.
See more detailed derivation and proofs in the Appendix.
\paragraph{Computational Complexity}
\vspace{-.2cm}Compared to those for rooted trees, the expected frequency counts $m_{s_1}^\mathrm{u}, m_{s,t}^\mathrm{u}, \overline{m}_{s_1}^\mathrm{u, (n)}, \overline{m}^{\mathrm{u},(n)}_{s,t}$ for unrooted trees involves additional summation over all the edges.
However, we can precompute all the accumulated sums of the root probabilities on the edges in a postorder tree traversal for each tree and the computation cost remains $\mathcal{O}(KN)$.
Each E-step in EM would in general cost $\mathcal{O}(KN^2)$ time since for each tree, $\mathcal{O}(N)$ probability estimations are required and each takes $\mathcal{O}(N)$ time.
Notice that most of the intermediate partial products involved in those SBN probability estimates are shared due to the structure of trees, we therefore use a two-pass algorithm, similar to the one used in \citep{Schadt1998}, that computes all SBN probability estimates for each tree within two loops over its edges.
This reduces the computational complexity of each E-step to $\mathcal{O}(KN)$.
Overall, the computational complexities of maximum lower bound estimate and each EM iteration are both $\mathcal{O}(KN)$, the same as CCD and SBNs for rooted trees.
In practice, EM usually takes several iterations to converge and hence could be more expensive than other methods.
However, the gain in approximation makes it a worthwhile trade-off (Table \ref{tab:realdata}).
We use the maximum SA lower bound algorithm ({\tt sbn-sa}), the EM algorithm ({\tt sbn-em}) and EM with regularization ({\tt sbn-em-$\alpha$}) in the experiment section.

\begin{figure}[t!]
\begin{center}
\includegraphics[width=\textwidth]{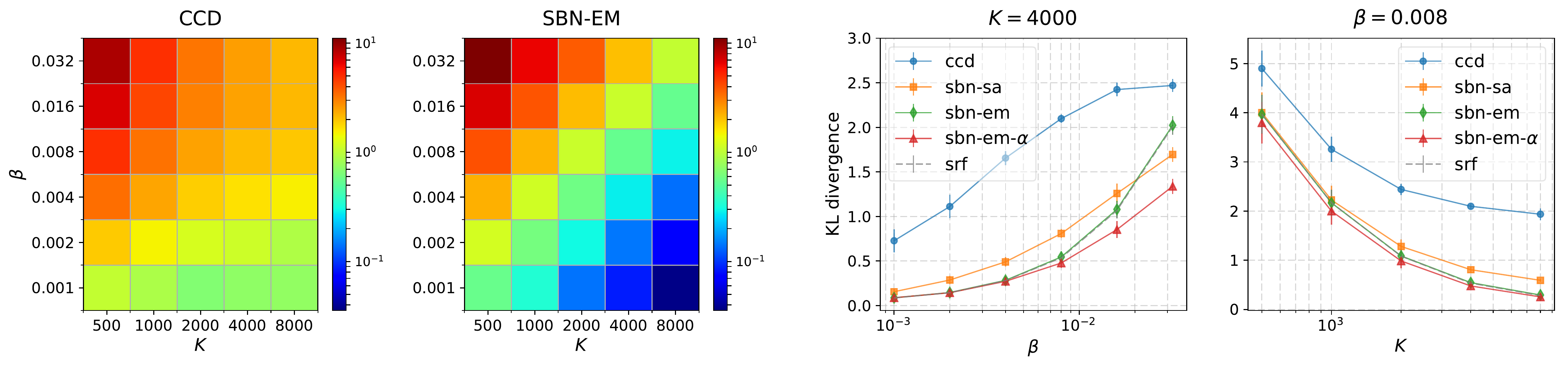}
\vskip -0.1in
\caption{Performance on a challenging tree probability estimation problem with simulated data.
{\bf Left:} The KL divergence of CCD and {\tt sbn-em} estimates over a wide range of degree of diffusion $\beta$ and sample size $K$.
{\bf Right:} A comparison among different methods for a fixed $K$, as a function of $\beta$ and for a fixed $\beta$, as a function of $K$.
Error bar shows one standard deviation over 10 runs.}\label{fig:simu}
\end{center}
\vskip-1em
\end{figure}
\section{Experiments}
\vspace{-0.3cm}We compare {\tt sbn-sa}, {\tt sbn-em}, {\tt sbn-em-$\alpha$} to the classical sample relative frequency (SRF) method and CCD on a synthetic data set and on estimating phylogenetic tree posteriors for a number of real data sets.
For all SBN algorithms, we use the simplest SBN, $\mathcal{B}_\mathcal{X}^\ast$, which we find provide sufficiently accurate approximation in the tree probability estimation tasks investigated in our experiments.
For {\tt sbn-em-$\alpha$}, we use the sample frequency counts of the root splits and parent-child subsplit pairs as the equivalent sample counts (see Algorithm~\ref{alg:SBNEM}).
The code is made available at {\tt https://github.com/zcrabbit/sbn}.

\paragraph{Simulated Scenarios}
\vspace{-0.3cm}To empirically explore the behavior of SBN algorithms relative to SRF and CCD, we first conduct experiments on a simulated setup.
We choose a tractable but challenging tree space, the space of unrooted trees with 8 leaves, which contains 10395 unique trees.
The trees are given an arbitrary order.
To test the approximation performance on targets of different degrees of diffusion, we generate target distributions by drawing samples from the Dirichlet distributions $\mathrm{Dir}(\beta {\mathbf 1})$ of order 10395 with a variety of $\beta$s.
The target distribution becomes more diffuse as $\beta$ increases.
Simulated data sets are then obtained by sampling from the unrooted tree space according to these target distributions with different sample sizes $K$.
The resulting probability estimation is challenging in that the target probabilities of the trees are assigned regardless of the similarity among them.
For {\tt sbn-em-$\alpha$}, we adjust the regularization coefficient using $\alpha=\frac{50}{K}$ for different sample sizes.
Since the target distributions are known, we use KL divergence from the estimated distributions to the target distributions to measure the approximation accuracy of different methods.
We vary $\beta$ and $K$ to control the difficulty of the learning task, and average over 10 independent runs for each configuration.
Figure \ref{fig:simu} shows the empirical approximation performance of different methods.
We see that the learning rate of CCD slows down very quickly as the data size increases, implying that the conditional clade independence assumption could be too strong to provide flexible approximations.
On the other hand, {\tt sbn-em} keeps learning efficiently from the data when more samples are available.
While all methods tend to perform worse as $\beta$ increases and perform better as $K$ increases, SBN algorithms performs consistently much better than CCD.
Compared to {\tt sbn-sa}, {\tt sbn-em} usually greatly improves the approximation with the price of additional computation.
When the degree of diffusion is large or the sample size is small, {\tt sbn-em-$\alpha$} gives much better performance than the others, showing that regularization indeed improves generalization.
See the Appendix for a runtime comparison of different methods with varying $K$ and $\beta$.

\paragraph{Real Data Phylogenetic Posterior Estimation}
\vspace{-0.3cm}We now investigate the performance on large unrooted tree space posterior estimation using 8 real datasets commonly used to benchmark phylogenetic MCMC methods \citep{Lakner08, Hhna2012-pm,Larget2013-et, Whidden2015-eq} (Table \ref{tab:realdata}).
For each of these data sets, 10 single-chain MrBayes \citep{Ronquist12} replicates are run for one billion iterations and sampled every 1000 iterations, using the simple \citet{Jukes69} substitution model.
We discard the first $25\%$ as burn-in for a total of 7.5 million posterior samples per data set.
These extremely long ``golden runs'' form the ground truth to which we will compare various posterior estimates based on standard runs.
For these standard runs, we run MrBayes on each data set with 10 replicates of 4 chains and 8 runs until the runs have ASDSF (the standard convergence criteria used in MrBayes) less than 0.01 or a maximum of 100 million iterations.
This conservative setting has been shown to find all posterior modes on these data sets \citep{Whidden2015-eq}.
We collect the posterior samples every 100 iterations of these runs and discard the first $25\%$ as burn-in.
We apply SBN algorithms, SRF and CCD to the posterior samples in each of the 10 replicates for each data set.
For {\tt sbn-em-$\alpha$}, we use $\alpha=0.0001$ to give some weak regularization\footnote{The same $\alpha$ is used for the real datasets since the sample sizes are roughly the same, although the number of unique trees are quite different.}.
We use KL divergence to the ground truth to measure the performance of all methods.

\begin{figure}[t!]
\begin{center}
\includegraphics[width=\textwidth]{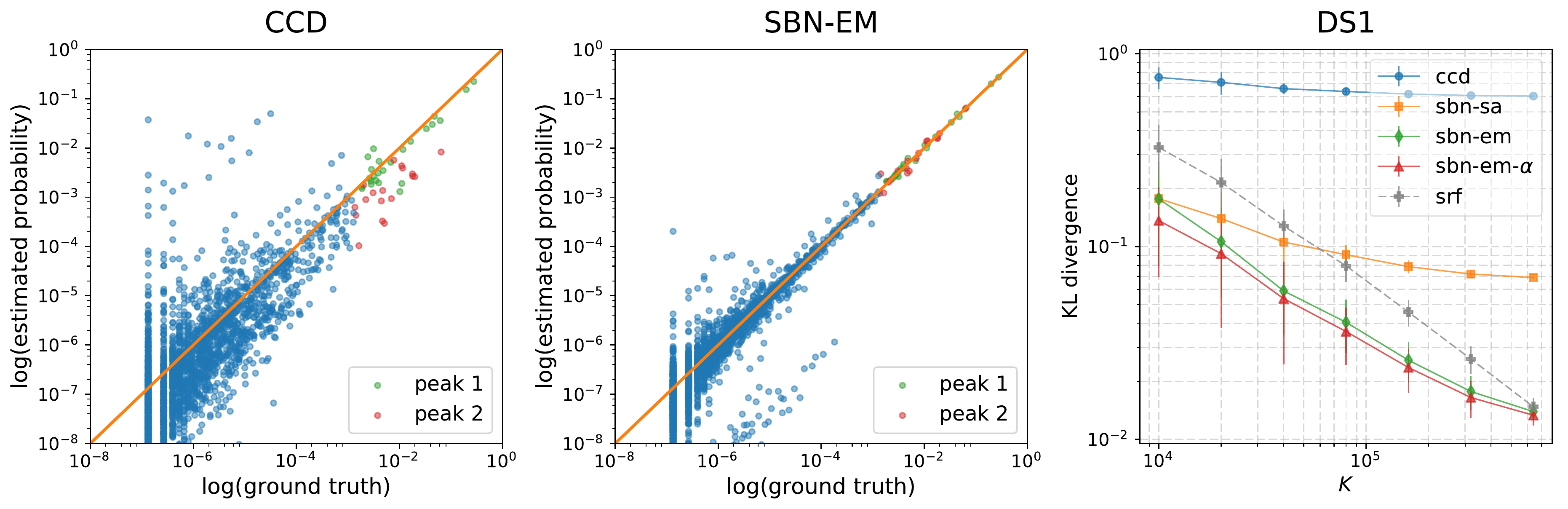}
\caption{Comparison on DS1, a data set with multiple posterior modes.
{\bf Left/Middle:} Ground truth posterior probabilities vs CCD and {\tt sbn-em} estimates.
{\bf Right:} Approximation performance as a function of sample size.
One standard deviation error bar over 10 replicates.}\label{fig:ds1}
\end{center}
\vskip -1.5em
\end{figure}

\begin{table}[t!]
\caption{Data sets used for phylogenetic posterior estimation, and approximation accuracy results of different methods across datasets.
Sampled trees column shows the numbers of unique trees in the standard run samples.
The results are averaged over 10 replicates.}
\label{tab:realdata}
\begin{center}
\begin{footnotesize}
\begin{sc}
\scalebox{0.65}{
\begin{tabular}{cccccccccc}
\toprule
\multirow{2}{*}{Data set} &\multirow{2}{*}{Reference} & \multirow{2}{*}{(\#Taxa, \#Sites)} & \multirow{2}{*}{\shortstack{Tree space\\ size}} &\multirow{2}{*}{\shortstack{Sampled\\trees}} &\multicolumn{5}{c}{KL divergence to ground truth} \\
\cmidrule(l){6-10}
  &     &    &     &   & SRF & CCD & SBN-SA & SBN-EM & SBN-EM-$\alpha$ \\
\midrule
DS1 & \citet{Hedges90}  & (27, 1949) & 5.84$\times$10\textsuperscript{32} & 1228 & 0.0155&0.6027&0.0687&0.0136&\bfseries 0.0130  \\
DS2 & \citet{Garey96} & (29, 2520) & 1.58$\times$10\textsuperscript{35}  & 7&\bfseries 0.0122&0.0218&0.0218&0.0199& 0.0128\\
DS3 & \citet{Yang03}  & (36, 1812) &4.89$\times$10\textsuperscript{47} & 43&0.3539&0.2074&0.1152&0.1243&\bfseries 0.0882\\
DS4 & \citet{Henk03} & (41, 1137) & 1.01$\times$10\textsuperscript{57} &828&0.5322&0.1952&0.1021&0.0763&\bfseries 0.0637 \\
DS5 & \citet{Lakner08} & (50, 378) & 2.84$\times$10\textsuperscript{74} & 33752 &11.5746 &1.3272 & 0.8952 & 0.8599 & \bfseries 0.8218 \\
DS6 & \citet{Zhang01} & (50, 1133) & 2.84$\times$10\textsuperscript{74} & 35407 &10.0159 & 0.4526 & \bfseries 0.2613 & 0.3016 & 0.2786 \\
DS7 & \citet{Yoder04}     & (59, 1824) & 4.36$\times$10\textsuperscript{92} &1125&1.2765&0.3292&0.2341&0.0483& \bfseries 0.0399\\
DS8 & \citet{Rossman01}   & (64, 1008) & 1.04$\times$10\textsuperscript{103} &3067&2.1653&0.4149&0.2212&0.1415& \bfseries 0.1236 \\
\bottomrule
\end{tabular}
}
\end{sc}
\end{footnotesize}
\end{center}
\vskip -0.2in
\end{table}

Previous work \citep{Whidden2015-eq} has observed that conditional clade independence does not hold in multimodal distributions.
Figure \ref{fig:ds1} shows a comparison on a typical data set, DS1, that has such a ``peaky'' distribution.
We see that CCD underestimates the probability of trees within the subpeak and overestimate the probability of trees between peaks.
In contrast, {\tt sbn-em} significantly removes these biases, especially for trees in the $95\%$ credible set.

When applied to a broad range of data sets, we find that SBNs consistently outperform other methods (Table \ref{tab:realdata}).
Due to its inability to generalize beyond observed samples, SRF is worse than generalizing probability estimators except for an exceedingly simple posterior with only 7 sampled trees (DS2).
CCD is, again, comparatively much worse than SBN algorithms.
With weak regularization, {\tt sbn-em-$\alpha$} gives the best performance in most cases.

To illustrate the efficiency of the algorithms on training data size, we perform an additional study on DS1 with increasing sample sizes and summarize the results in the right panel of Figure \ref{fig:ds1}.
As before, we see that CCD slows down quickly while SBN algorithms, especially fully-capable SBN estimators {\tt sbn-em} and {\tt sbn-em-$\alpha$}, keep learning efficiently as the sample size increases.
Moreover, SBN algorithms tend to provide much better approximation than SRF when fewer samples are available, which is important in practice where large samples are expensive to obtain.

\section{Conclusion}
\vspace{-0.3cm}We have proposed a general framework for tree probability estimation based on subsplit Bayesian networks.
SBNs allow us to exploit the similarity among trees to provide a wide range of flexible probability estimators that generalize beyond observations.
Moreover, they also allow many efficient Bayesian network learning algorithms to be extended to tree probability estimation with ease.
We report promising numerical results demonstrating the importance of being both flexible and generalizing when estimating probabilities on trees.
Although we present SBNs in the context of leaf-labeled bifurcating trees, it can be easily adapted for general leaf-labeled trees by allowing partitions other than subsplits (bipartitions) of the clades in parent nodes.
We leave for future work investigating the performance of more complicated SBNs for general trees, structure learning of SBNs, deeper examination of the effect of parameter sharing, and further applications of SBNs to other probabilistic learning problems in tree spaces, such as designing more efficient tree proposals for MCMC transition kernels and providing flexible and tractable distributions for variational inference.

\section*{Acknowledgements}
This work supported by National Science Foundation grant CISE-1564137, as well as National Institutes of Health grants R01-GM113246 and U54-GM111274.
The research of Frederick Matsen was supported in part by a Faculty Scholar grant from the Howard Hughes Medical Institute and the Simons Foundation.

\bibliography{sbn}
\bibliographystyle{plainnat}

\appendix

\section{The SBN Representation of The Rooted Tree from Figure \ref{fig:tree}}
\begin{figure}[!h]
\begin{center}
\input{figs/bn-example}
\caption{The SBN representation of the rooted tree from Figure \ref{fig:tree}. For ease of presentation, we use the indexes of the leaf labels for subsplits. We also omit the fake splits after depth 4.}
\end{center}
\end{figure}
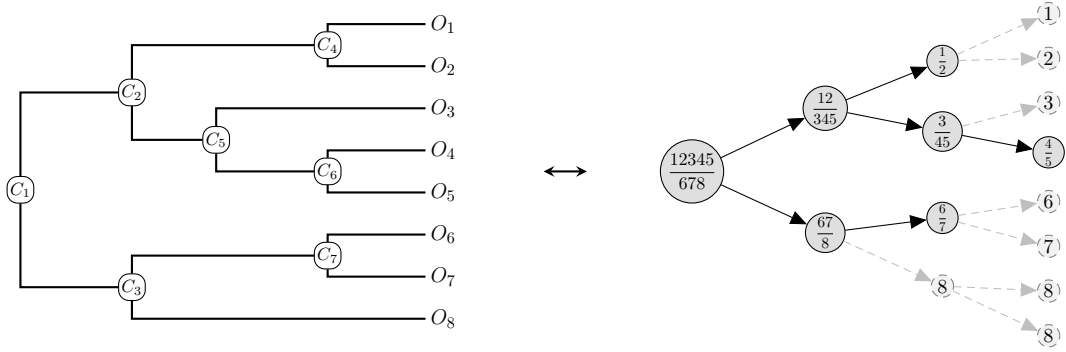

\section{Proof of Lemma \ref{lemma:sbn}}
\begin{proof}
It suffices to prove the lemma for the minimum SBN $\mathcal{B}_\mathcal{X}^\ast$. For any rooted tree $T$, with natural indexing, a compatible assignment is easy to obtain by following the splitting process and satisfying the compatibility requirement defined in Definition \ref{def:compatible}. Moreover, the depth of the splitting process is at most $N-1$, where $N$ is the size of $\mathcal{X}$. To show this, consider the maximum size of clades, $d_i$, in the $i$-th level of the process. The first level is for the root split, and $d_1 \le N-1$ since the subclades need to be proper subset of $\mathcal{X}$. For each $i>1$, if $d_i>1$, the splitting process continues and the maximum size of clades in the next level is at most $d_i-1$, that is $d_{i+1}\leq d_i-1$. Note that the split process ends when $d_i=1$, its depth therefore is at most $N-1$. The whole process is deterministic, so each rooted tree can be uniquely represented as a compatible assignment of $\mathcal{B}_\mathcal{X}^\ast$. On the other hand, given a compatible assignment, we can simply remove all fake splits. The remaining net corresponds to the true splitting process of a rooted tree, which then can be simply restored.
\end{proof}

\section{Proof of Proposition \ref{prop:rooted}}
\begin{proof}
Denote the set of all SBN assignments as $\mathcal{A}$. For any assignment $a = \{S_i=s_i^a\}_{i\ge 1}$, we have
\begin{equation}
p_{\mathrm{sbn}}(a) = p(S_1=s_1^a) \prod_{i>1}p(S_i=s_i^a|S_{\pi_i} = s_{\pi}^a) > 0 \Rightarrow \text{ $a$ is compatible}.
\end{equation}
As a Bayesian network,
\[
\sum_{a\in \mathcal{A}}p_{\mathrm{sbn}}(a) = \sum_{s_1}p(S_1=s_1)\prod_{i>1}\sum_{s_i}p(S_i=s_i|S_{\pi_i}=s_{\pi_i}) = 1.
\]
By Lemma \ref{lemma:sbn},
\[
\sum_{T}p_{\mathrm{sbn}}(T) = \sum_{a \sim {\mathrm{compatible}}} p_{\mathrm{sbn}}(a) = \sum_{a\in \mathcal{A}}p_{\mathrm{sbn}}(a) = 1.
\]
\end{proof}

\section{Figure for Shared Parent-child Subsplit Pairs in SBNs}
\begin{figure}[h!]
\begin{center}
\input{figs/cp-sharing}
\caption{Subsplit pairs shared among different nodes in SBNs.
Top panels show the rooted trees that share the same local parent-child subsplit pair $((B,C),D)$.
Bottom panels are their corresponding representations in SBNs.
Just like in the trees, the same pair could appear at multiple locations in SBNs.}\label{fig:CPS}
\end{center}
\end{figure}
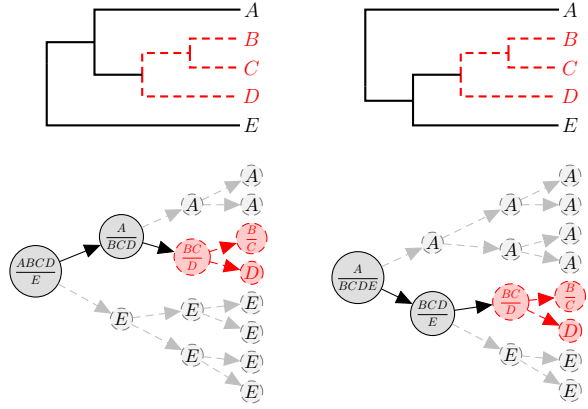

\section{Proof of Proposition \ref{prop:unrooted}}
\begin{proof}
\begin{align*}
\sum_{T^\mathrm{u}} p_{\mathrm{sbn}}(T^\mathrm{u}) &= \sum_{T^\mathrm{u}} \sum_{S_1\sim T^{\mathrm{u}}}p(S_1)\prod_{i>1}p(S_i|S_{\pi_i})\\
&= \sum_{T}p_{\mathrm{sbn}}(T)\\
&= 1.
\end{align*}
The second equality is due to the one-to-one correspondence between rooted trees and unrooted trees with specific rooting edges (compatible splits):
\[
T \overset{\mathrm{1\text{-}to\text{-}1}}{\Longleftrightarrow} (T^\mathrm{u}, S_1), \; S_1 \sim T^{\mathrm{u}}.
\]
\end{proof}

\section{On Expectation Maximization for SBNs}
Using the conditional probabilities of the missing root node, we can construct the following adaptive lower bound
\begin{equation}\label{eq:emlb}
LB^{(n)}(T^\mathrm{u};p) = \sum_{S_1\sim T^\mathrm{u}} p(S_1|T^\mathrm{u}, \hat{p}^{\mathrm{EM},(n)}) \log \left( \dfrac{p(S_1) \prod_{i>1} p(S_i|S_{\pi_i})}{p(S_1|T^\mathrm{u}, \hat{p}^{\mathrm{EM},(n)})}\right) \leq \log L(T^\mathrm{u};p).
\end{equation}
The above adaptive lower bound $LB^{(n)}(T^\mathrm{u};p)$ contains a constant term that only depends on the current estimates, and another term that is the expected complete log-likelihood with respect to the conditional distribution of $S_1$ given $T^\mathrm{u}$ and the current estimates:
\begin{equation}\label{eq:estep}
Q^{(n)}(T^\mathrm{u};p) = \sum_{S_1\sim T^\mathrm{u}} p(S_1|T^\mathrm{u}, \hat{p}^{\mathrm{EM},(n)})\left(\log p(S_1) + \sum_{i>1} \log p(S_i|S_{\pi_i})\right).
\end{equation}

Summing \eqref{eq:estep} over $T^\mathrm{u}\in \mathcal{D}^\mathrm{u}$ together with conditional probability sharing, we get the complete data score function
\begin{equation}\label{eq:score}
\begin{aligned}
Q^{(n)}(\mathcal{D}^\mathrm{u};p) &= \sum_{k=1}^K Q^{(n)}(T_k^\mathrm{u};p)\\
&= \sum_{k=1}^K\sum_{e\in E(T_k^\mathrm{u})} q_k^{(n)}(S_1=s_{1,k}^e)\Big(\log p(S_1=s_{1,k}^e) + \sum_{i>1}\log p(S_i=s_{i,k}^e|S_{\pi_i}=s_{\pi_i,k}^e)\Big)\\
&= \sum_{s_1\in\mathbb{C}_r}\overline{m}_{s_1}^\mathrm{u, (n)}  \log p(S_1=s_1) + \sum_{s|t\in\mathbb{C}_{\mathrm{ch}|\mathrm{pa}}} \overline{m}_{s,t}^\mathrm{u, (n)} \log p(s|t)
\end{aligned}
\end{equation}
where
\[
\overline{m}_{s_1}^\mathrm{u, (n)}=\sum_{k=1}^K\sum_{e\in E(T_k^\mathrm{u})} q_k^{(n)}(S_1=s_1) \mathbbold{I}({s_{i,k}^e=s_1})
\]
\[
\overline{m}^{\mathrm{u},(n)}_{s,t} = \sum_{k=1}^K\sum_{e\in E(T_k^{\mathrm{u}})}q_k^{(n)}(S_1=s_{1,k}^e)\sum_{i>1} \mathbbold{I}({s_{i,k}^e=s, s_{\pi_i,k}^e = t})
\]
are the expected frequency counts.

\subsection{Proof of Theorem \ref{thm:em}}
\begin{proof}
\begin{align*}
Q^{(n)}(T^\mathrm{u};p) - Q^{(n)}(T^\mathrm{u}; \hat{p}^{\mathrm{EM},(n)}) &= LB^{(n)}(T^\mathrm{u};p) - LB^{(n)}(T^\mathrm{u}; \hat{p}^{\mathrm{EM},(n)}) \\
&\leq \log L(T^\mathrm{u};p) - \log L(T^\mathrm{u}; \hat{p}^{\mathrm{EM},(n)})
\end{align*}
since $LB^{(n)}(T^\mathrm{u};p) \leq \log L(T^\mathrm{u};p)$ and
\begin{align*}
LB^{(n)}(T^\mathrm{u}; \hat{p}^{\mathrm{EM},(n)}) &= \sum_{S_1\sim T^\mathrm{u}} p(S_1|T^\mathrm{u}, \hat{p}^{\mathrm{EM},(n)}) \log \left( \dfrac{p(T^\mathrm{u}, S_1|\hat{p}^{\mathrm{EM},(n)})}{p(S_1|T^\mathrm{u}, \hat{p}^{\mathrm{EM},(n)})}\right)\\
&= \sum_{S_1\sim T^\mathrm{u}} p(S_1|T^\mathrm{u}, \hat{p}^{\mathrm{EM},(n)}) \log p(T^\mathrm{u}|\hat{p}^{\mathrm{EM},(n)})\\
&= \log p(T^\mathrm{u}|\hat{p}^{\mathrm{EM},(n)})\\
&= \log \sum_{S_1\sim T^\mathrm{u}} p(T^\mathrm{u},S_1|\hat{p}^{\mathrm{EM},(n)})\\
&= \log L(T^\mathrm{u}; \hat{p}^{\mathrm{EM},(n)}).
\end{align*}
\end{proof}

With Theorem \ref{thm:em}, it is clear that
\[
Q^{(n)}(\mathcal{D}^\mathrm{u};p) - Q^{(n)}(\mathcal{D}^\mathrm{u}; \hat{p}^{\mathrm{EM},(n)})  \le  \log L(\mathcal{D}^\mathrm{u};p) - \log L(\mathcal{D}^\mathrm{u}; \hat{p}^{\mathrm{EM},(n)}).
\]

As before, the maximum estimates of $Q^{(n)}$ have closed form expressions which lead to the following updating formula for the CPDs at the M-step
\begin{equation}\label{eq:emroot}
\hat{p}^{\mathrm{EM},(n+1)}(S_1=s_1) = \frac{\overline{m}_{s_1}^{\mathrm{u},(n)}}{\sum_{s\in\mathbb{C}_r} \overline{m}_s^{\mathrm{u},(n)}} = \frac{\overline{m}_{s_1}^{\mathrm{u},(n)}}{K}, \quad s_1\in\mathbb{C}_r
\end{equation}
\begin{equation}\label{eq:emchild}
\hat{p}^{\mathrm{EM},(n+1)}(s|t) = \frac{\overline{m}_{s,t}^{\mathrm{u},(n)}}{\sum_s \overline{m}_{s,t}^{\mathrm{u},(n)}},\quad s|t \in \mathbb{C}_{\mathrm{ch}|\mathrm{pa}}.
\end{equation}

\subsection{Regularization}
For the score function in \eqref{eq:score}, the conjugate prior distributions have to be $\operatorname{Dirichlet}$ distributions
\[
S_1 \sim \mathrm{Dir}(\alpha_{s_1}+1, s_1\in \mathbb{C}_r),\quad \cdot |t \sim \mathrm{Dir}(\alpha_{s,t}+1, s|t\in\mathbb{C}_{\mathrm{ch}|\mathrm{pa}})
\]
where $\alpha_{s_1}>0,\;\alpha_{s,t}>0$ are some hyper-parameters. The regularized score function, therefore, takes the following form
\begin{equation}
\begin{aligned}
Q^{\mathrm{R}, (n)}(\mathcal{D}^\mathrm{u};p) &= Q^{(n)}(\mathcal{D}^\mathrm{u};p) + \sum_{s_1\in\mathbb{C}_r}\alpha_{s_1}\log p(S_1=s_1) + \sum_{s|t\in\mathbb{C}_{\mathrm{ch}}|\mathrm{pa}} \alpha_{s,t}\log p(s|t)\\
&=\sum_{s_1\in\mathbb{C}_r}(\overline{m}_{s_1}^\mathrm{u, (n)} + \alpha_{s_1}) \log p(S_1=s_1) + \sum_{s|t\in\mathbb{C}_{\mathrm{ch}|\mathrm{pa}}} (\overline{m}_{s,t}^\mathrm{u, (n)} +\alpha_{s,t})\log p(s|t).
\end{aligned}
\end{equation}
One can adapt the hyper-parameters according to this decomposition as follows
\[
\alpha_{s_1} = \alpha\cdot \tilde{m}_{s_1}^\mathrm{u},\quad \alpha_{s,t} = \alpha \cdot\tilde{m}_{s,t}^\mathrm{u}
\]
where $\alpha$ is the global regularization coefficient that balances the effect of the data on the estimates in comparison to the prior, and $\tilde{m}_{s_1}^\mathrm{u}, \tilde{m}_{s,t}^\mathrm{u}$ are the equivalent sample counts for the root splits and parent-child subsplit pairs that distribute the regularization across the CPDs. In practice, one can simply use the the sample frequency counts as equivalent sample counts as we did in our experiments.

{\bf Remark: } The choice of hyper-parameter setting is not throughly studied and what we do here is some modification from the common practice in Bayesian estimation for learning Bayesian networks; see \citet{Buntine91} for more details.

Theorem \ref{thm:em} also implies that maximizing (or improving) the regularized $Q$ score is sufficient to improve the regularized log-likelihood. Denote the regularization term as
\[
R(p) = \sum_{s_1\in\mathbb{C}_r}\alpha\tilde{m}_{s_1}^\mathrm{u}\log p(S_1=s_1) + \sum_{s|t\in\mathbb{C}_{\mathrm{ch}}|\mathrm{pa}} \alpha\tilde{m}_{s,t}^\mathrm{u}\log p(s|t)
\]
we have
\begin{align*}
Q^{\mathrm{R}, (n)}(\mathcal{D}^\mathrm{u};p) - Q^{\mathrm{R}, (n)}(\mathcal{D}^\mathrm{u}; \hat{p}^{\mathrm{EM},(n)}) &= Q^{(n)}(\mathcal{D}^\mathrm{u};p) - Q^{(n)}(\mathcal{D}^\mathrm{u}; \hat{p}^{\mathrm{EM},(n)}) + R(p) - R(\hat{p}^{\mathrm{EM},(n)})\\
&\le \log L(\mathcal{D}^\mathrm{u};p) + R(p) - \big(\log L(\mathcal{D}^\mathrm{u}; \hat{p}^{\mathrm{EM},(n)}) + R(\hat{p}^{\mathrm{EM},(n)})\big).
\end{align*}

Similarly, the regularized score $Q^{\mathrm{R}, (n)}$ can be maximized in the same manner as $Q^{(n)}$
\begin{equation}\label{eq:emroot}
\hat{p}^{\mathrm{EM},(n+1)}(S_1=s_1) = \frac{\overline{m}_{s_1}^{\mathrm{u},(n)}+\alpha\tilde{m}_{s_1}^\mathrm{u}}{\sum_{s_1\in\mathbb{C}_r} (\overline{m}_{s_1}^{\mathrm{u},(n)}+\alpha\tilde{m}_{s_1}^\mathrm{u})} = \frac{\overline{m}_{s_1}^{\mathrm{u},(n)}+\alpha\tilde{m}_{s_1}^\mathrm{u}}{K+\alpha\sum_{s_1\in\mathbb{C}_r}\tilde{m}_{s_1}^\mathrm{u}}, \quad s_1\in\mathbb{C}_r
\end{equation}
\begin{equation}\label{eq:emchild}
\hat{p}^{\mathrm{EM},(n+1)}(s|t) = \frac{\overline{m}_{s,t}^{\mathrm{u},(n)}+\alpha\tilde{m}_{s,t}^\mathrm{u}}{\sum_s (\overline{m}_{s,t}^{\mathrm{u},(n)}+\alpha\tilde{m}_{s,t}^\mathrm{u})},\quad s|t \in \mathbb{C}_{\mathrm{ch}|\mathrm{pa}}.
\end{equation}

\section{Run Time Comparison}
\begin{figure}[h!]
\begin{center}
\includegraphics[width=\textwidth]{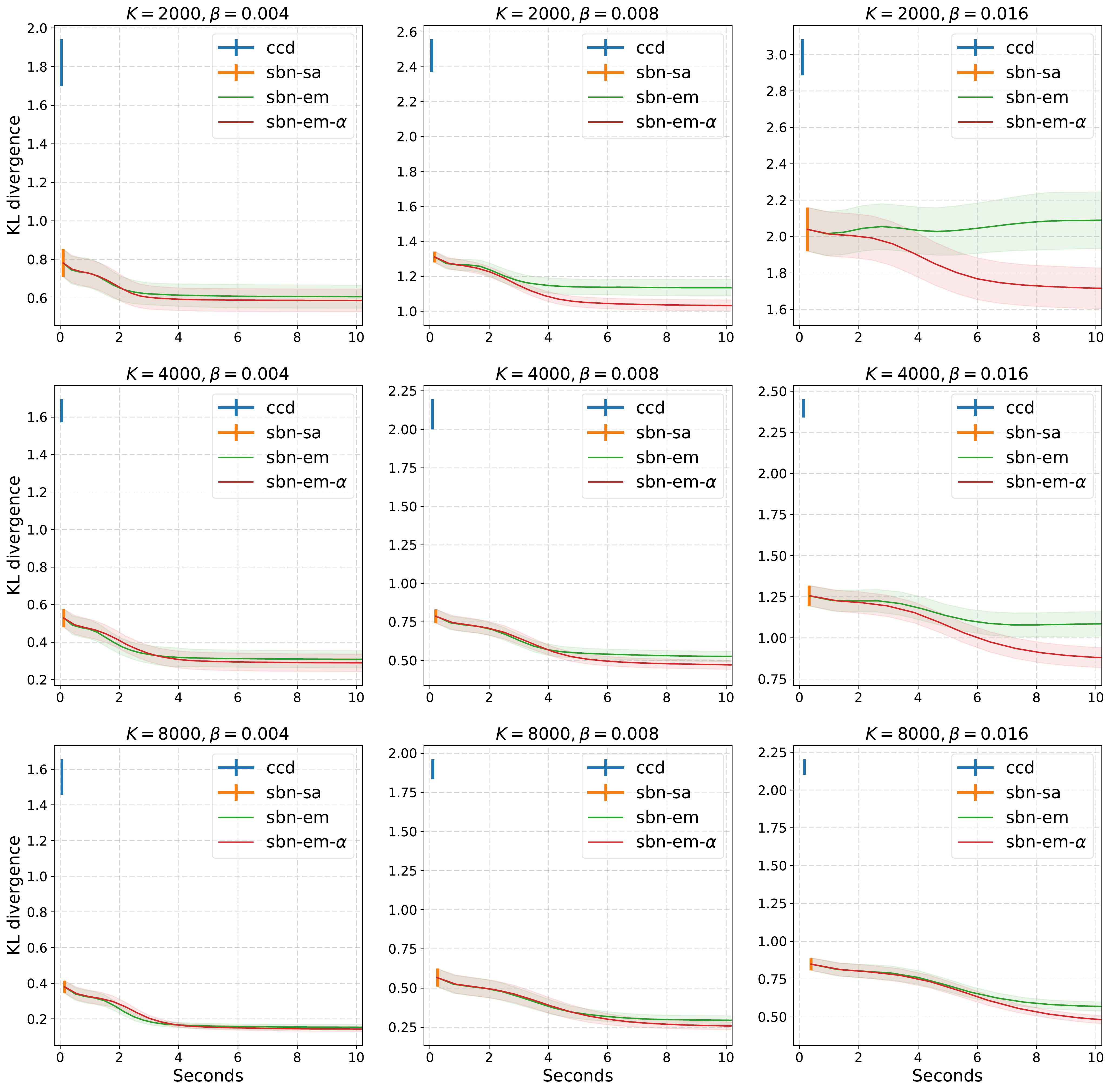}
\caption{A runtime comparison of different methods with varying $K$ and $\beta$ on simulated data. Error bar shows one standard deviation over 10 runs.}\label{fig:runtime}
\end{center}
\end{figure}

In this section, we present a runtime comparison of different algorithms on simulated data with varying $K$ and $\beta$. We see that SBN based methods significantly improve the approximation accuracy of CCD. With additional time budget, {\tt sbn-em} and  {\tt sbn-em-$\alpha$} further improve the performance of {\tt sbn-sa}, and regularization is useful especially for diffuse distributions with limited sample size (Figure~\ref{fig:runtime}). All experiments were done on a 2016 MacBook Pro (2.9 GHz Intel Core i5).
%

\end{document}

%% file: figs/clade-decomp-square-edge.tex
\input{rooted_tree}

\tikz[scale=0.48, transform shape] {
  \begin{scope}[rounded corners]
  \node[leaf, xshift=7cm, yshift=3.5cm] (A) {$O_1$};
  \node[leaf, xshift=7cm, yshift=2.5cm] (B) {$O_2$};
  \node[leaf, xshift=7cm, yshift=1.5cm] (C) {$O_3$};
  \node[leaf, xshift=7cm, yshift=0.5cm] (D) {$O_4$};
  \node[leaf, xshift=7cm, yshift=-0.5cm] (E) {$O_5$};
  \node[leaf, xshift=7cm, yshift=-1.5cm] (F) {$O_6$};
  \node[leaf, xshift=7cm, yshift=-2.5cm] (G) {$O_7$};
  \node[leaf, xshift=7cm, yshift=-3.5cm] (H) {$O_8$};
  \node[clade, left=of A, yshift=-0.5cm] (C4) {\scalebox{1.2}{$C_4$}};
  \node[clade, left=of D, yshift=-0.5cm] (C6) {\scalebox{1.2}{$C_6$}};
  \node[clade, left=of F, yshift=-0.5cm] (C7) {\scalebox{1.2}{$C_7$}};
  \node[clade, left=of C4, xshift=-2cm, yshift=-1.125cm] (C2) {\scalebox{1.2}{$C_2$}};
  \node[clade, left=of C6, yshift=0.75cm] (C5) {\scalebox{1.2}{$C_5$}};
  \node[clade, left=of C7, xshift=-2cm, yshift=-0.75cm] (C3) {\scalebox{1.2}{$C_3$}};
  \node[clade, left=of C2, yshift=-2.3125cm] (C1) {\scalebox{1.2}{$C_1$}};
  \end{scope}
  \squaredge[thick] {C1}{C2, C3};
  \squaredge[thick] {C2}{C4, C5};
  \squaredge[thick] {C4}{A, B};
  \squaredge[thick] {C5}{C, C6};
  \squaredge[thick] {C6}{D, E};
  \squaredge[thick] {C3}{C7, H};
  \squaredge[thick] {C7}{F, G};

}

%% file: figs/bn-reversed.tex
\input{rooted_tree}
\tikz[scale=0.65]{
  \begin{scope}[xscale=0.52, yscale=0.52, transform shape]
  \node[obs](S1){\scalebox{1.8}{$S_1$}};
  \node[obs, right=of S1, xshift=0.75cm, yshift=1.8cm](S2){\scalebox{1.8}{$S_2$}};
  \node[obs, right=of S1, xshift=0.75cm, yshift=-1.8cm](S3){\scalebox{1.8}{$S_3$}};
  \node[obs, right=of S2, xshift=0.75cm, yshift=1.8cm](S4){\scalebox{1.8}{$S_4$}};
  \node[obs, right=of S2, xshift=0.75cm, yshift=-0.6cm](S5){\scalebox{1.8}{$S_5$}};
  \node[obs, right=of S3, xshift=0.75cm, yshift=0.6cm](S6){\scalebox{1.8}{$S_6$}};
  \node[obs, right=of S3, xshift=0.75cm, yshift=-1.8cm](S7){\scalebox{1.8}{$S_7$}};
  \node[obs, right=of S4, xshift=0.75cm, yshift=1.8cm](S8){\scalebox{1.8}{$\cdots$}};
  \node[obs, right=of S4, xshift=0.75cm, yshift=0.25cm](S9){\scalebox{1.8}{$\cdots$}};
  \node[obs, right=of S5, xshift=0.75cm, yshift=1.05cm](S10){\scalebox{1.8}{$\cdots$}};
  \node[obs, right=of S5, xshift=0.75cm, yshift=-0.45cm](S11){\scalebox{1.8}{$\cdots$}};
  \node[obs, right=of S6, xshift=0.75cm, yshift=0.45cm](S12){\scalebox{1.8}{$\cdots$}};
  \node[obs, right=of S6, xshift=0.75cm, yshift=-1.05cm](S13){\scalebox{1.8}{$\cdots$}};
  \node[obs, right=of S7, xshift=0.75cm, yshift=-0.25cm](S14){\scalebox{1.8}{$\cdots$}};
  \node[obs, right=of S7, xshift=0.75cm, yshift=-1.8cm](S15){\scalebox{1.8}{$\cdots$}};
  \edge{S1}{S2, S3} ;
  \edge{S2}{S4, S5} ;
  \edge{S3}{S6, S7} ;
  \edge{S4}{S8, S9};
  \edge{S5}{S10, S11};
  \edge{S6}{S12, S13};
  \edge{S7}{S14, S15};
  \path (S1) edge [->, >={triangle 45}, bend right=10, dashed]  (S5);
  \path (S1) edge [->, >={triangle 45}, bend right=-30, dashed]  (S4);
  \path (S1) edge [->, >={triangle 45}, bend right=-10, dashed]  (S6);
  \path (S1) edge [->, >={triangle 45}, bend right=30, dashed]  (S7);
  \path (S2) edge [->, >={triangle 45}, bend right=5, dashed]  (S9);
  \path (S2) edge [->, >={triangle 45}, bend right=-30, dashed]  (S8);
  \path (S2) edge [->, >={triangle 45}, bend right=-5, dashed]  (S10);
  \path (S2) edge [->, >={triangle 45}, bend right=20, dashed]  (S11);
  \path (S3) edge [->, >={triangle 45}, bend right=-20, dashed]  (S12);
  \path (S3) edge [->, >={triangle 45}, bend right=5, dashed]  (S13);
  \path (S3) edge [->, >={triangle 45}, bend right=-5, dashed]  (S14);
  \path (S3) edge [->, >={triangle 45}, bend right=20, dashed]  (S15);
  \end{scope}
  
  \begin{scope}[xscale=0.52, yscale=0.52, transform shape, xshift=16cm, yshift=-1cm]
  \node[obs, yshift=4cm] (T1SP1) {\scalebox{1.2}{$\dfrac{ABC}D$}};

  \node[obs, right=of T1SP1, xshift=0.5cm, yshift=1.1cm](T1SP2){\scalebox{1.2}{$\dfrac{A}{BC}$}};
  \node[obs, right=of T1SP1, xshift=0.7cm, yshift=-1.1cm, fill=gray!10, draw=gray, densely dashed](T1SP3){\scalebox{1.8}{$D$}};
  \node[obs, right=of T1SP2, xshift=0.5cm, yshift=1.1cm, fill=gray!10, draw=gray, densely dashed](T1SP4) {\scalebox{1.8}{$A$}};
  \node[obs, right=of T1SP2, xshift=0.4cm, yshift=-0.35cm](T1SP5) {\scalebox{1.2}{$\dfrac{B}C$}};
  \node[obs, right=of T1SP3, xshift=.75cm, yshift=0.35cm, fill=gray!10, draw=gray, densely dashed] (T1SP6) {\scalebox{1.8}{$D$}};
  \node[obs, right=of T1SP3, xshift=.75cm, yshift=-1.1cm, fill=gray!10, draw=gray, densely dashed] (T1SP7) {\scalebox{1.8}{$D$}};
    
  \edge{T1SP1}{T1SP2} 
  \annotatededge[densely dashed, gray!50]{pos=0.3, below=1pt}{1.0}{T1SP1}{T1SP3};
  \annotatededge[densely dashed, gray!50]{pos=0.3, above=2pt}{1.0}{T1SP2}{T1SP4};
  \edge{T1SP2}{T1SP5};
  \annotatededge[densely dashed, gray!50]{pos=0.3, above=2pt}{1.0}{T1SP3}{T1SP6};
  \annotatededge[densely dashed, gray!50]{pos=0.3, below=1pt}{1.0}{T1SP3}{T1SP7};

  \node[obs, yshift=-2.5cm] (T2SP1) {\scalebox{1.2}{$\dfrac{AB}{CD}$}};
  \node[obs, right=of T2SP1, xshift=0.5cm, yshift=1.1cm] (T2SP2) {\scalebox{1.2}{$\dfrac{A}{B}$}};
  \node[obs, right=of T2SP1, xshift=0.5cm, yshift=-1.1cm] (T2SP3) {\scalebox{1.2}{$\dfrac{C}{D}$}};
  \node[obs, right=of T2SP2, xshift=0.75cm, yshift=1.1cm, fill=gray!10, draw=gray, densely dashed] (T2SP4) {\scalebox{1.8}{$A$}};
  \node[obs, right=of T2SP2, xshift=0.75cm, yshift=-0.4cm, fill=gray!10, draw=gray, densely dashed] (T2SP5) {\scalebox{1.8}{$B$}};
  \node[obs, right=of T2SP3, xshift=0.75cm, yshift=0.4cm, fill=gray!10, draw=gray, densely dashed] (T2SP6) {\scalebox{1.8}{$C$}};
  \node[obs, right=of T2SP3, xshift=0.75cm, yshift=-1.1cm, fill=gray!10, draw=gray, densely dashed] (T2SP7) {\scalebox{1.8}{$D$}}; 
  
  \edge{T2SP1}{T2SP2, T2SP3};
  \annotatededge[densely dashed, gray!50]{pos=0.3, above=2pt}{1.0}{T2SP2}{T2SP4};
  \annotatededge[densely dashed, gray!50]{pos=0.3, below=1pt}{1.0}{T2SP2}{T2SP5};
  \annotatededge[densely dashed, gray!50]{pos=0.3, above=2pt}{1.0}{T2SP3}{T2SP6};
  \annotatededge[densely dashed, gray!50]{pos=0.3, below=1pt}{1.0}{T2SP3}{T2SP7};
  
  \end{scope}   

  \begin{scope}[xscale=0.52, yscale=0.52, transform shape, yshift=-1cm, xshift=30cm]
  \node[yshift=2.2cm] (T1S1) {};
  \node[left=2.5cm of T1S1, xshift=-2cm, yshift=1.6cm] (T1Arrow) {};
  \draw[<->, >=stealth, thick] (T1Arrow) -- +(1.5,0);
  \node[right=of T1S1, xshift=0.5cm, yshift=3.2cm](T1S2) {};
  \node[leaf, right=of T1S1, xshift=3.48cm] (T1D) {\scalebox{1.4}{$D$}};
  \node[leaf, right=of T1S2, xshift=1.8cm, yshift=0.75cm] (T1A) {\scalebox{1.4}{$A$}};
  \node[right=of T1S2, xshift=0.5cm, yshift=-1.2cm] (T1S3) {};
  \node[leaf, right=of T1S3, yshift=0.6cm] (T1B) {\scalebox{1.4}{$B$}};
  \node[leaf, right=of T1S3, yshift=-0.6cm] (T1C) {\scalebox{1.4}{$C$}};
  \csquaredge[thick]{T1S1}{T1S2}
  \csquaredgeleaf[thick] {T1S1}{T1D};
   \csquaredgeleaf[thick] {T1S2}{T1A};
  \csquaredge[thick]{T1S2}{T1S3};
  \csquaredgeleaf[thick]{T1S3}{T1B, T1C};

  \node[yshift=-2.5cm] (T2S1) {};
  \node[left=2.5cm of T2S1, xshift=-2cm] (T2Arrow) {};
  \draw[<->, >=stealth, thick] (T2Arrow) -- +(1.5, 0);
  \node[right=of T2S1, xshift=0.75cm, yshift=1.25cm] (T2S2) {};
  \node[right=of T2S1, xshift=0.75cm, yshift=-1.25cm] (T2S3) {};
  \node[leaf, right=of T2S2, xshift=1.45cm, yshift=0.75cm] (T2A) {\scalebox{1.4}{$A$}};
  \node[leaf, right=of T2S2, xshift=1.45cm, yshift=-0.6cm] (T2B) {\scalebox{1.4}{$B$}};
  \node[leaf, right=of T2S3, xshift=1.45cm, yshift=0.6cm] (T2C) {\scalebox{1.4}{$C$}};
  \node[leaf, right=of T2S3, xshift=1.45cm, yshift=-0.75cm] (T2D) {\scalebox{1.4}{$D$}};
  \csquaredge[thick]{T2S1}{T2S2, T2S3};
  \csquaredgeleaf[thick]{T2S2}{T2A, T2B};
  \csquaredgeleaf[thick]{T2S3}{T2C, T2D};
  \end{scope}
  
%
%
}

%% file: figs/bn-unrooted.tex
\input{rooted_tree}
\tikzstyle{fakenode} = [obs, fill=gray!10, draw=gray, densely dashed]
\tikz[scale=0.75]{
  \begin{scope}[xscale=0.45, yscale=0.45, transform shape]
%
  \node[xshift=1.5cm, yshift=-2.5cm] (UTUp) {};
  \node[below=of UTUp, yshift=-0.25cm] (UTDown) {};
  \node[leaf, above=of UTUp, xshift=-1.5cm, yshift=-0.75cm] (UTA) {\scalebox{1.4}{$A$}};
  \node[leaf, above=of UTUp, xshift=1.5cm, yshift=-0.75cm] (UTB) {\scalebox{1.4}{$B$}};
  \node[leaf, below=of UTDown, xshift=-1.5cm, yshift=0.75cm] (UTC) {\scalebox{1.4}{$C$}};
  \node[leaf, below=of UTDown, xshift=1.5cm, yshift=0.75cm] (UTD) {\scalebox{1.4}{$D$}}; 
  \undiredgeleaf[thick]{pos=0.6, below=1pt, font=\Large}{1}{UTUp}{UTA}
  \undiredgeleaf[thick]{pos=0.6, below=1pt,font=\Large}{2}{UTUp}{UTB};
  \undiredgeleaf[thick]{pos=0.6, above=2pt,font=\Large}{4}{UTDown}{UTC};
  \undiredgeleaf[thick]{pos=0.6, above=2pt,font=\Large}{5}{UTDown}{UTD};
  \undiredge[thick]{pos=0.5, right=1pt,font=\Large}{3}{UTUp}{UTDown};
  \end{scope}
  
  \begin{scope}[xscale=0.45, yscale=0.45, transform shape, xshift=9cm]
  \node[leaf, xshift=5cm, yshift=2cm] (T1A){\scalebox{1.4}{$A$}};
  \node[leaf, below=of T1A, yshift=1.3cm] (T1B) {\scalebox{1.4}{$B$}};
  \node[leaf, below=of T1B, yshift=1.3cm] (T1C) {\scalebox{1.4}{$C$}};
  \node[leaf, below=of T1C, yshift=1.3cm] (T1D) {\scalebox{1.4}{$D$}};
  \node[left = of T1C, xshift=-0.6cm, yshift=-0.7cm] (T1S3) {};
  \node[left = of T1S3, xshift=0.5cm, yshift=1.05cm] (T1S2) {};
  \node[left = of T1S2, xshift=0.5cm, yshift=1.15cm] (T1S1) {};
  \node[left = of T1S1, xshift=0.75cm, font=\Large] (T1root) {$1$};
  
  \node[left=of T1root, xshift=-2.5cm, yshift=-1.1cm] (T1Arrow) {};
  \path (T1Arrow) edge[<->, >= stealth, thick] node[pos=0.5, above=4pt, font=\LARGE, sloped]{root/unroot} +(1.3, 0.75);
  
  \csquaredgeleaf[thick] {T1S3} {T1C, T1D};
  \csquaredgeleaf[thick] {T1S2} {T1B};
  \csquaredge[thick] {T1S2}{T1S3};
  \csquaredgeleaf[thick] {T1S1} {T1A};
  \csquaredge[thick] {T1S1} {T1S2};
  \draw[thick] (T1root) -- (T1S1.center);  
  

  \node[leaf, below=of T1D, yshift=0.8cm] (T2A){\scalebox{1.4}{$A$}};
  \node[leaf, below=of T2A, yshift=1.3cm] (T2B) {\scalebox{1.4}{$B$}};
  \node[leaf, below=of T2B, yshift=1.3cm] (T2C) {\scalebox{1.4}{$C$}};
  \node[leaf, below=of T2C, yshift=1.3cm] (T2D) {\scalebox{1.4}{$D$}};
  \node[left = of T2A, xshift=-1.4cm, yshift=-0.7cm] (T2S3) {};
  \node[left = of T2C, xshift=-1.4cm, yshift=-0.7cm] (T2S2) {};
  \node[left = of T2S2, xshift=0.5cm, yshift=1.25cm] (T2S1) {};
  \node[left = of T2S1, xshift=0.75cm, font=\Large] (T2root) {$3$};
  
  \node[left=of T2root, xshift=-2.5cm, yshift=0.8cm] (T2Arrow) {};
  \path (T2Arrow) edge[<->, >=stealth, thick] node[pos=0.5, below=1pt, font=\LARGE, sloped]{root/unroot} +(1.3,-0.75);
  
  \csquaredgeleaf[thick]{T2S3}{T2A, T2B};
  \csquaredgeleaf[thick]{T2S2}{T2C, T2D};
  \csquaredge[thick]{T2S1}{T2S2, T2S3};
  \draw[thick] (T2root) -- (T2S1.center);
  \end{scope}
  
  \begin{scope}[xscale=0.45, yscale=0.45, transform shape, xshift=20cm]
  \node[fakenode, xshift=6cm, yshift=2cm] (N1S4){\scalebox{1.8}{$A$}};
  \node[fakenode, xshift=6cm, yshift=0.6cm] (N1S5) {\scalebox{1.8}{$A$}};
  \node[fakenode, xshift=6cm, yshift=-0.8cm] (N1S6) {\scalebox{1.8}{$B$}};
  \node[obs, xshift=6cm, yshift=-2.2cm] (N1S7) {\scalebox{1.2}{$\dfrac{C}{D}$}}; 
  
 \node[fakenode, left= of N1S4, xshift=-0.5cm, yshift=-1cm] (N1S2) {\scalebox{1.8}{$A$}};
 \node[obs, below = of N1S2, yshift=-0.12cm] (N1S3) {\scalebox{1.2}{$\dfrac{B}{CD}$}};
 \node[obs, left= of N1S2, xshift=-0.2cm, yshift=-1cm] (N1S1) {\scalebox{1.2}{$\dfrac{A}{BCD}$}};
 
 \node[left=of N1S1, xshift=-2.5cm] (N1Arrow) {};
 \path (N1Arrow) edge[<->, >=stealth, thick] +(1.5,0);
 
 \annotatededge[densely dashed, gray!50]{pos=0.4, above=2pt}{}{N1S2}{N1S4, N1S5};
 \annotatededge[densely dashed, gray!50]{pos=0.4, above=2pt}{}{N1S3}{N1S6};
 \edge{N1S3}{N1S7};
 \annotatededge[densely dashed, gray!50]{pos=0.4, above=2pt}{}{N1S1}{N1S2};
 \edge{N1S1}{N1S3};
  
  \node[fakenode, xshift=6cm, yshift=-4.4cm] (N2S4){\scalebox{1.8}{$A$}};
  \node[fakenode, xshift=6cm, yshift=-5.8cm] (N2S5) {\scalebox{1.8}{$B$}};
  \node[fakenode, xshift=6cm, yshift=-7.2cm] (N2S6) {\scalebox{1.8}{$C$}};
  \node[fakenode, xshift=6cm, yshift=-8.6cm] (N2S7) {\scalebox{1.8}{$D$}}; 
  
 \node[obs, left= of N2S4, xshift=-0.5cm, yshift=-1cm] (N2S2) {\scalebox{1.2}{$\dfrac{A}{B}$}};
 \node[obs, below = of N2S2, yshift=-0.08cm] (N2S3) {\scalebox{1.2}{$\dfrac{C}{D}$}};
 \node[obs, left= of N2S2, xshift=-0.2cm, yshift=-1cm] (N2S1) {\scalebox{1.2}{$\dfrac{AB}{CD}$}};
 
  \node[left=of N2S1, xshift=-2.5cm] (N2Arrow) {};
 \path (N2Arrow) edge[<->, >=stealth, thick] +(1.5,0);
 
 \annotatededge[densely dashed, gray!50]{pos=0.4, above=2pt}{}{N2S2}{N2S4, N2S5};
 \annotatededge[densely dashed, gray!50]{pos=0.4, above=2pt}{}{N2S3}{N2S6, N2S7};
 \edge{N2S1}{N2S2, N2S3};
 
 \node[right=of N1S5, xshift=1.75cm, yshift=-0.7cm] (N1Arrowout) {};
 \path (N1Arrowout) edge[<->, >=stealth, thick] +(1.3,-0.75);
 \node[right=of N2S6, xshift=1.75cm, yshift=0.7cm] (N2Arrowout) {};
 \path (N2Arrowout) edge[<->, >=stealth, thick] +(1.3,0.75);
 \end{scope}
 \begin{scope}[xscale=0.6, yscale=0.6, transform shape, xshift=25.5cm, yshift=.5cm]
  \node[obs, xshift=4cm, yshift=0cm] (UNS4) {\scalebox{1.6}{$S_4$}};
  \node[obs, xshift=4cm, yshift=-2cm] (UNS5) {\scalebox{1.6}{$S_5$}};
  \node[obs, xshift=4cm, yshift=-4cm] (UNS6) {\scalebox{1.6}{$S_6$}};
  \node[obs, xshift=4cm, yshift=-6cm] (UNS7) {\scalebox{1.6}{$S_7$}};
  
  \node[obs, left=of UNS4, xshift=-0.5cm, yshift=-1.6cm] (UNS2) {\scalebox{1.6}{$S_2$}};
  \node[obs, below=of UNS2, yshift=-0.95cm] (UNS3) {\scalebox{1.6}{$S_3$}};
  \node[latent, left=of UNS2, xshift=-0.2cm, yshift=-1.4cm] (UNS1) {\scalebox{1.6}{$S_1$}};
  
  \edge {UNS2}{UNS4, UNS5};
  \edge {UNS3}{UNS6, UNS7};
  \edge {UNS1}{UNS2, UNS3};
 \end{scope}
}

%% file: figs/bn-example.tex
\input{rooted_tree}
\tikzstyle{fakenode} = [obs, fill=gray!10, draw=gray, densely dashed]
\tikz[scale=0.7]{
  \begin{scope}[xscale=0.8, yscale=0.8, transform shape, rounded corners]
  \node[leaf, xshift=7cm, yshift=3.5cm] (A) {$O_1$};
  \node[leaf, xshift=7cm, yshift=2.5cm] (B) {$O_2$};
  \node[leaf, xshift=7cm, yshift=1.5cm] (C) {$O_3$};
  \node[leaf, xshift=7cm, yshift=0.5cm] (D) {$O_4$};
  \node[leaf, xshift=7cm, yshift=-0.5cm] (E) {$O_5$};
  \node[leaf, xshift=7cm, yshift=-1.5cm] (F) {$O_6$};
  \node[leaf, xshift=7cm, yshift=-2.5cm] (G) {$O_7$};
  \node[leaf, xshift=7cm, yshift=-3.5cm] (H) {$O_8$};
  \node[clade, left=of A, yshift=-0.5cm] (C4) {\scalebox{1.2}{$C_4$}};
  \node[clade, left=of D, yshift=-0.5cm] (C6) {\scalebox{1.2}{$C_6$}};
  \node[clade, left=of F, yshift=-0.5cm] (C7) {\scalebox{1.2}{$C_7$}};
  \node[clade, left=of C4, xshift=-2cm, yshift=-1.125cm] (C2) {\scalebox{1.2}{$C_2$}};
  \node[clade, left=of C6, yshift=0.75cm] (C5) {\scalebox{1.2}{$C_5$}};
  \node[clade, left=of C7, xshift=-2cm, yshift=-0.75cm] (C3) {\scalebox{1.2}{$C_3$}};
  \node[clade, left=of C2, yshift=-2.3125cm] (C1) {\scalebox{1.2}{$C_1$}};
  \end{scope}
  \squaredge[thick] {C1}{C2, C3};
  \squaredge[thick] {C2}{C4, C5};
  \squaredge[thick] {C4}{A, B};
  \squaredge[thick] {C5}{C, C6};
  \squaredge[thick] {C6}{D, E};
  \squaredge[thick] {C3}{C7, H};
  \squaredge[thick] {C7}{F, G};
    
  \begin{scope}[xscale=0.6, yscale=0.6, transform shape, xshift=23.5cm, yshift=5cm]
    \node[fakenode, xshift=5cm] (SP8) {\scalebox{1.8}{$1$}};
    \node[fakenode, below=of SP8, yshift=0.3cm] (SP9) {\scalebox{1.8}{$2$}};
    \node[obs, left=of SP8, xshift=-1.5cm, yshift=-1.5cm] (SP4) {\scalebox{1.2}{$\dfrac{1}{2}$}};
    \node[obs, below=of SP4, yshift=-0.1cm] (SP5) {\scalebox{1.4}{$\dfrac{3}{45}$}};
    \node[fakenode, below=of SP9, yshift=0.3cm] (SP10) {\scalebox{1.8}{$3$}};
    \node[obs, below=of SP10, yshift=0.3cm] (SP11) {\scalebox{1.2}{$\dfrac{4}{5}$}};
    \node[fakenode, below=of SP11, yshift=0.3cm] (SP12) {\scalebox{1.8}{$6$}};
    \node[fakenode, below=of SP12, yshift=0.3cm] (SP13) {\scalebox{1.8}{$7$}};
    \node[obs, below=of SP5, yshift=-0.6cm] (SP6){\scalebox{1.2}{$\dfrac{6}{7}$}};
    \node[fakenode, below=of SP6, yshift=-0.3cm] (SP7) {\scalebox{1.8}{$8$}};
    \node[fakenode, below=of SP13, yshift=0.3cm] (SP14) {\scalebox{1.8}{$8$}};
    \node[fakenode, below=of SP14, yshift=0.3cm] (SP15) {\scalebox{1.8}{$8$}};
    \node[obs, left=of SP4, xshift=-1.5cm, yshift=-1.5cm] (SP2) {\scalebox{1.4}{$\dfrac{12}{345}$}};
    \node[obs, below=of SP2, yshift=-1.6cm] (SP3) {\scalebox{1.4}{$\dfrac{67}{8}$}};
    \node[obs, left=of SP2, xshift=-1.5cm, yshift=-2.0cm] (SP1) {\scalebox{1.6}{$\dfrac{12345}{678}$}};
    
    \annotatededge[densely dashed, gray!50] {}{}{SP4}{SP8, SP9};
    \annotatededge[densely dashed, gray!50] {}{}{SP5}{SP10};
    \edge{SP5}{SP11};
    \annotatededge[densely dashed, gray!50] {}{}{SP6}{SP12, SP13};
    \annotatededge[densely dashed, gray!50] {}{}{SP7}{SP14, SP15};
    \edge{SP2}{SP4, SP5};
    \edge{SP3}{SP6};
    \annotatededge[densely dashed, gray!50] {}{}{SP3}{SP7};
    \edge{SP1}{SP2, SP3};
  \end{scope}
  
  \node[left=0.5cm of SP1, xshift=-1.05cm] (Arrow) {};
  \draw[<->, >=stealth, thick] (Arrow) -- + (1, 0);
  }

%% file: figs/cp-sharing.tex
\input{rooted_tree}
\tikzstyle{fakenode} = [obs, fill=gray!10, draw=gray, densely dashed]
\tikz[scale=0.7]{
\begin{scope}[xscale=0.55, yscale=0.55, transform shape]
  \node[leaf, xshift=7cm, yshift=2cm](T1A) {\scalebox{1.4}{$A$}};
  \node[leaf, xshift=7cm, yshift=1cm, red](T1B) {\scalebox{1.4}{$B$}};
  \node[leaf, xshift=7cm, yshift=0cm, red](T1C) {\scalebox{1.4}{$C$}};
  \node[leaf, xshift=7cm, yshift=-1cm, red](T1D) {\scalebox{1.4}{$D$}};
  \node[leaf, xshift=7cm, yshift=-2cm](T1E) {\scalebox{1.4}{$E$}};
  \node[left=of T1B, xshift=-0.6cm, yshift=-0.5cm] (T1S4) {};
  \node[left=of T1S4, xshift=-0.4cm, yshift=-0.75cm] (T1S3) {};
  \node[left=of T1S3, xshift=-0.4cm, yshift=1.125cm] (T1S2) {};
  \node[left=of T1S2, xshift=-0.4cm, yshift=-0.75cm] (T1S1) {};
    
  \csquaredgeleaf[thick,red, densely dashed]{T1S4}{T1B,T1C};
  \csquaredge[thick,red, densely dashed]{T1S3}{T1S4};
  \csquaredgeleaf[thick, red, densely dashed]{T1S3}{T1D};
  \csquaredge[thick]{T1S2}{T1S3};
  \csquaredgeleaf[thick]{T1S2}{T1A};
  \csquaredge[thick]{T1S1}{T1S2};
  \csquaredgeleaf[thick]{T1S1}{T1E};
  
  \node[fakenode, below=of T1E, yshift=0cm](N1S8){\scalebox{1.8}{$A$}};
  \node[fakenode, below=of N1S8, yshift=0.75cm](N1S9){\scalebox{1.8}{$A$}};
  \node[obs, below=of N1S9, yshift=0.75cm, draw=red, fill=red!20, densely dashed](N1S10){\scalebox{1.2}{$\textcolor{red}{\dfrac{B}{C}}$}};
  \node[obs, below=of N1S10, yshift=0.75cm, draw=red, fill=red!20, densely dashed](N1S11){\scalebox{1.8}{$\textcolor{red}{D}$}};
  \node[fakenode, below=of N1S11, yshift=0.75cm](N1S12){\scalebox{1.8}{$E$}};
  \node[fakenode, below=of N1S12, yshift=0.75cm](N1S13){\scalebox{1.8}{$E$}};
  \node[fakenode, below=of N1S13, yshift=0.75cm](N1S14){\scalebox{1.8}{$E$}};
  \node[fakenode, below=of N1S14, yshift=0.75cm](N1S15){\scalebox{1.8}{$E$}};
  \node[fakenode, left=of N1S8, xshift=-0.3cm, yshift=-1cm] (N1S4){\scalebox{1.8}{$A$}};
  \node[obs, below=of N1S4, yshift=0.2cm, draw=red, fill=red!20, densely dashed] (N1S5) {\scalebox{1.2}{$\textcolor{red}{\dfrac{BC}{D}}$}};
  \node[fakenode, below=of N1S5, yshift=0.2cm] (N1S6) {\scalebox{1.8}{$E$}};
  \node[fakenode, below=of N1S6, yshift=0.2cm] (N1S7) {\scalebox{1.8}{$E$}};
  \node[obs, left=of N1S4, xshift=-0.3cm, yshift=-1.1cm] (N1S2) {\scalebox{1.2}{$\dfrac{A}{BCD}$}};
  \node[fakenode, below=of N1S2, yshift=-0.7cm] (N1S3) {\scalebox{1.8}{$E$}};
  \node[obs, left=of N1S2, xshift=-0.3cm, yshift=-1.2cm](N1S1){\scalebox{1.2}{$\dfrac{ABCD}{E}$}};
  
  \annotatededge[densely dashed, gray!50]{pos=0.4, above=2pt}{}{N1S4}{N1S8};
  \annotatededge[densely dashed, gray!50]{pos=0.4, below=1pt}{}{N1S4}{N1S9};
  \edge[red, densely dashed]{N1S5}{N1S10, N1S11};
  \annotatededge[densely dashed, gray!50]{pos=0.4, above=2pt}{}{N1S6}{N1S12, N1S13};
  \annotatededge[densely dashed, gray!50]{pos=0.4, below=1pt}{}{N1S7}{N1S14, N1S15};
  \annotatededge[densely dashed, gray!50]{pos=0.4, above=2pt}{}{N1S2}{N1S4};
  \edge{N1S2}{N1S5};
  \annotatededge[densely dashed, gray!50]{pos=0.4, above=2pt}{}{N1S3}{N1S6, N1S7};
  \edge{N1S1}{N1S2};
  \annotatededge[densely dashed, gray!50]{pos=0.4, below=1pt}{}{N1S1}{N1S3};

\end{scope}

\begin{scope}[xscale=0.55, yscale=0.55, transform shape, xshift=10cm]
 \node[leaf, xshift=8cm, yshift=2cm](T1A) {\scalebox{1.4}{$A$}};
  \node[leaf, xshift=8cm, yshift=1cm, red](T1B) {\scalebox{1.4}{$B$}};
  \node[leaf, xshift=8cm, yshift=0cm, red](T1C) {\scalebox{1.4}{$C$}};
  \node[leaf, xshift=8cm, yshift=-1cm, red](T1D) {\scalebox{1.4}{$D$}};
  \node[leaf, xshift=8cm, yshift=-2cm](T1E) {\scalebox{1.4}{$E$}};
  \node[left=of T1B, xshift=-0.6cm, yshift=-0.5cm] (T1S4) {};
  \node[left=of T1S4, xshift=-0.4cm, yshift=-0.75cm] (T1S3) {};
  \node[left=of T1S3, xshift=-0.4cm, yshift=-0.9cm] (T1S2) {};
  \node[left=of T1S2, xshift=-0.4cm, yshift=0.5cm] (T1S1) {};
    
  \csquaredgeleaf[thick,red, densely dashed]{T1S4}{T1B,T1C};
  \csquaredge[thick,red, densely dashed]{T1S3}{T1S4};
  \csquaredgeleaf[thick, red, densely dashed]{T1S3}{T1D};
  \csquaredge[thick]{T1S2}{T1S3};
  \csquaredgeleaf[thick]{T1S2}{T1E};
  \csquaredge[thick]{T1S1}{T1S2};
  \csquaredgeleaf[thick]{T1S1}{T1A};
  
  \node[fakenode, below=of T1E, yshift=0cm](N1S8){\scalebox{1.8}{$A$}};
  \node[fakenode, below=of N1S8, yshift=0.75cm](N1S9){\scalebox{1.8}{$A$}};
  \node[fakenode, below=of N1S9, yshift=0.75cm](N1S10){\scalebox{1.8}{$A$}};
  \node[fakenode, below=of N1S10, yshift=0.75cm](N1S11){\scalebox{1.8}{$A$}};
   \node[obs, below=of N1S11, yshift=0.75cm, draw=red, fill=red!20, densely dashed](N1S12){\scalebox{1.2}{$\textcolor{red}{\dfrac{B}{C}}$}};
  \node[obs, below=of N1S12, yshift=0.75cm, draw=red, fill=red!20, densely dashed](N1S13){\scalebox{1.8}{$\textcolor{red}{D}$}};
  \node[fakenode, below=of N1S13, yshift=0.75cm](N1S14){\scalebox{1.8}{$E$}};
  \node[fakenode, below=of N1S14, yshift=0.75cm](N1S15){\scalebox{1.8}{$E$}};
  \node[fakenode, left=of N1S8, xshift=-0.3cm, yshift=-1cm] (N1S4){\scalebox{1.8}{$A$}};
  \node[fakenode, below=of N1S4, yshift=0.2cm] (N1S5) {\scalebox{1.8}{$A$}};
  \node[obs, below=of N1S5, yshift=0.2cm, draw=red, fill=red!20, densely dashed] (N1S6) {\scalebox{1.2}{$\textcolor{red}{\dfrac{BC}{D}}$}};
  \node[fakenode, below=of N1S6, yshift=0.2cm] (N1S7) {\scalebox{1.8}{$E$}};
  \node[obs, left=of N1S6, xshift=-0.3cm, yshift=-0.4cm] (N1S3) {\scalebox{1.2}{$\dfrac{BCD}{E}$}};
  \node[fakenode, above=of N1S3, yshift=0.3cm] (N1S2) {\scalebox{1.8}{$A$}};
  \node[obs, left=of N1S2, xshift=-0.3cm, yshift=-1.2cm](N1S1){\scalebox{1.2}{$\dfrac{A}{BCDE}$}};
  
  \annotatededge[densely dashed, gray!50]{pos=0.4, above=2pt}{}{N1S4}{N1S8};
  \annotatededge[densely dashed, gray!50]{pos=0.4, below=1pt}{}{N1S4}{N1S9};
  \edge[red, densely dashed]{N1S6}{N1S12, N1S13};
  \annotatededge[densely dashed, gray!50]{pos=0.4, above=2pt}{}{N1S5}{N1S10, N1S11};
  \annotatededge[densely dashed, gray!50]{pos=0.4, below=1pt}{}{N1S7}{N1S14, N1S15};
  \annotatededge[densely dashed, gray!50]{pos=0.4, above=2pt}{}{N1S2}{N1S4, N1S5};
  \edge{N1S3}{N1S6};
  \annotatededge[densely dashed, gray!50]{pos=0.4, above=2pt}{}{N1S3}{N1S7};
  \edge{N1S1}{N1S3};
  \annotatededge[densely dashed, gray!50]{pos=0.4, below=1pt}{}{N1S1}{N1S2};
\end{scope}


}